%% file: arxiv_draft.tex
\documentclass[11pt]{article}
\usepackage{todonotes}
\input{_header}

\usepackage{hyperref}
\usepackage{float}
\usepackage{threeparttable}

\title{Participation and Representation\\in Local Government Speech}
\author{Olivia Martin \and Amar Venugopal}
\date{This version: April 22, 2026}
\begin{document}

\maketitle

\begin{abstract}

Local government meetings are the most common formal channel through which residents speak directly with elected officials, contest policies, and shape local agendas. However, data constraints typically limit the empirical study of these meetings to agendas, single cities, or short time horizons. We collect and transcribe a massive new dataset of city council meetings from 115 California cities over the last decade, using advanced transcription and diarization techniques to analyze the speech content of the meetings themselves. We document two sets of descriptive findings: First, city council meetings are frequent, long, and vary modestly across towns and time in topical content. Second, public participants are substantially older, whiter, more male, more liberal, and more likely to own homes than the registered voter population, and public participation surges when topics related to land use and zoning are included in meeting agendas. Given this skew, we examine the main policy lever municipalities have to shift participation patterns: meeting access costs. Exploiting pandemic-era variation in remote access, we show that eliminating remote options reduces the number of speakers, but does not clearly change the composition of speakers. Collectively, these results provide the most comprehensive empirical portrait to date of who participates in local democracy, what draws them in, and how institutional design choices shape both the volume and composition of public input.

\end{abstract}

\nnfootnote{Authors are listed in alphabetical order. Olivia Martin (\href{mailto:omartin@stanford.edu}{omartin@stanford.edu}), Department of Economics, Stanford University; Amar Venugopal (\href{mailto:amarvenu@stanford.edu}{amarvenu@stanford.edu}), Department of Economics, Stanford University.  We thank Diego Zambrano, Jann Spiess, Guido Imbens, Daniel E. Ho, Matt Gentzkow, Mark Duggan, Nick Bloom, Alexia Olaizola, Peter Robertson, and members of the Stanford Econometrics workshop for helpful comments and suggestions. We thank Ellen Kwon and Samay Rele for excellent research assistance. We thank the Stanford Law School Neukom Center for the Rule of Law, the Stanford Institute for Human-Centered AI (HAI) and the John M. Olin Program in Law and Economics for their generous support. All errors are our own.}

\pagebreak
\pagenumbering{arabic}
\setcounter{page}{1}

\section{Introduction}
City governments are economically important institutions. They employ millions of workers, spend hundreds of billions of dollars each year, and make decisions over land use, public safety, infrastructure, parks, and other place-based services that shape household welfare and local economic activity \citep{us_census_bureau_2022_2024}. In the classic fiscal federalism view \citep{oates_essay_1999}, these are precisely the kinds of policy domains in which decentralized decision-making can be valuable, because local officials are better positioned to respond to heterogeneous local conditions and preferences. The efficacy of these institutions is predicated on the ability of local governments to effectively aggregate preferences of the local population. In this paper, we focus on the most direct and persistent formal forum for citizen participation in local government: city council meetings.

Under California’s Brown Act and analogous open-meeting laws nationwide, local elected bodies must deliberate in public, post agendas in advance, and provide opportunities for residents to speak. Yet despite the scale of the decisions made in these settings, data constraints have resulted in typically limited empirical evidence on the dynamics and characteristics of local government meetings, with existing datasets either relying on agendas or minutes and therefore failing to capture the actual text of discussion (e.g., \citet{einstein_who_2019}) or collecting full transcripts but restricting attention to a potentially unrepresentative sample of cities (e.g., \citet{barari_localview_2023}). As a consequence, we know surprisingly little about who actually participates, what kinds of issues draw them in, and how institutional design affects the costs of participation, factors which are central to local governments' ability to effectively aggregate local preferences.

To answer these questions, we construct a new dataset of over 25,000 city council meetings across 115 California municipalities from 2015–2024, comprising over 75,000 hours of audio generated by over 150,000 meeting participants that we transcribe, diarize, and annotate using state-of-the-art language models. We further augment these data via linking to voter registration and property ownership records. With this newly constructed dataset, we are able to provide the largest, most comprehensive characterization of meeting structure, participation, and content to date.

Our analysis proceeds in three steps. First, we provide a comprehensive characterization of city council meetings in California. Meetings are long, frequent, and have many participants. From the text of the meeting transcripts, we further identify issues of discussion and associated vote outcomes, thereby recovering meeting agendas. We then classify these issues into distinct meta-topics of interest to city councils. We find that most council votes are unanimous and that cities exhibit modest variation in topics of discussion. 

Second, we conduct a comprehensive study of meeting participation, and find significant representation gaps between participants and registered voters: participants tend to be older, less likely to be female, more likely to be Democrats, more likely to be of White non-Hispanic origin, and more likely to be a homeowner than the registered voter population. These same demographic characteristics also drive selection into repeat participation. Participants also exhibit differential tendencies to appear in a meeting based on the topic mix of issues listed on the meeting agenda, with meetings emphasizing ``land use and zoning" seeing substantially more public participation. We further find that while city-level characteristics impact participation rates, representation gaps in participants persist. As a consequence of these gaps, the participant pool is systematically different from the registered voter pool. This speaks to a core concern regarding the efficacy of local governance: if the participant pool is unrepresentative, then city governments are unable to effectively aggregate local preferences. 

Third, we present a simple framework for an individual’s participation decision, in which a registered voter enters the participant pool if the benefits of participation (in terms of discussion of relevant content) outweighs the costs. Under this framework, there are two potential policy levers available to local governments to increase representativeness: they may either alter the topical content of their meetings, or they may alter meeting access costs. Since local governments are typically constrained in what topics they must discuss, we focus on changing access costs as the key policy lever available to local governments to improve representation. By leveraging quasi-experimental variation in meeting access costs during and after the COVID-19 pandemic, we are able to estimate causal effects of changing meeting access costs on participant representativeness. We find that increasing meeting access costs leads to lower participation rates and alters the age makeup of the participant pool.

Taken together, our results constitute the most comprehensive empirical study to date of the mechanics of local democracy, allowing us to provide novel characterizations of how meetings are structured, what is discussed, who participates, and how both topics of discussion and variable costs of meeting access drive trends in participation.

\section{Related Literature}
This paper contributes to several literatures on public economics, political economy, political science, and computational social science. 

First, our paper relates to work in public economics, urban economics, and local political economy on decentralized governance and the politics of local public goods. In classic fiscal-federalism accounts, decentralization can improve welfare when local governments are better able to respond to heterogeneous local preferences and conditions \citep{oates_essay_1999, besley_centralized_2003}. A large related literature emphasizes that local institutions are shaped by unequal political participation, differential stakes, and the distribution of property ownership \citep{fischel_homevoter_2009, tausanovitch_representation_2014, warshaw_local_2019}. These concerns are especially salient in land use and housing, where local political voice can be highly asymmetric and where homeowners often have both stronger incentives and greater capacity to participate \citep{fischel_homevoter_2009, yoder_does_2020, einstein_who_2019, hankinson_when_2018, marble_where_2021}. Our paper speaks to this literature by documenting participation gaps in one central local institution at scale and investigating the efficacy of a convenient policy lever available to local governments to reduce these gaps.

Second, our paper contributes to a growing empirical literature in political science on public meetings, hearings, and descriptive representation in local politics. Scholars have emphasized that municipal institutions are understudied relative to their importance in structuring daily life and public service provision \citep{trounstine_representation_2010, warshaw_local_2019}. Related work has also documented representational skew in other local participatory settings, including land-use hearings and planning processes, and has begun to connect commenter preferences to policy outcomes in particular agencies and cities \citep{sahn_public_2025, einstein_still_2023}.

The closest substantive antecedent to our work in this strand of the literature is \citet{einstein_who_2019}, who show that participants in suburban planning and zoning meetings are disproportionately older, whiter, male, and homeowner-heavy relative to the broader public. We extend on their analysis in several ways. First, our dataset exhibits far greater geographic and temporal breadth, covering 115 diverse cities in California over the last decade, compared to 97 towns in metropolitan Boston in a three-year span. Consequently, we observe a far greater number of public participants who we are able match to voter records (over 100,000, compared to approximately 3,000), and collect full transcripts rather than minutes. Additionally, we collect data on city council meetings, a materially different venue than planning and zoning boards which covers a far wider range of topics. Finally, by collecting data before, during, and after the COVID-19 pandemic, we are able to speak to the variable effect of remote participation options on meeting structure, content, and participation. 

Third, our paper also relates to work on participation costs, institutional design, and the accessibility of democratic fora. A long tradition in political economy treats participation as sensitive to the costs of time, travel, scheduling, and information \citep{downs_economic_1957, riker_theory_1968}. Those cost considerations are especially important in local settings, where meetings are often held on weekday evenings, at fixed physical locations, and on agendas that vary substantially in salience. Recent work on local meetings during and after the pandemic suggests that remote and hybrid formats can alter who appears in these fora, though the effects on representation are not straightforward \citep{einstein_still_2023}. Our contribution is to bring a larger panel and a cleaner institutional design to this question. Leveraging staggered withdrawals of remote public participation across cities, we study how lowering or raising the cost of attendance affects both the extensive margin of participation and the demographic composition of speakers, investigating a key policy lever available to local governments to address documented representation gaps.

Finally, our paper contributes to the computational social science and text-as-data literature by showing how modern language models can convert a large, messy corpus of institutional speech into analyzable data. Existing work in computational social science has developed tools for measuring meaning and group differences in political and administrative text \citep{gentzkow_measuring_2019, card_computational_2022}. More recent work has used large language models for classification, extraction, and annotation tasks in economics and related fields, including the analysis of regulatory and legal texts \citep{bartik_costs_2023, modarressi_causal_2025, durvasula_counting_2025}. The closest large-scale data precedent in the local-government context is the \textit{LocalView} project, which assembled a large corpus of local government meeting transcripts from YouTube and demonstrated the promise of these materials for social-scientific research \citep{barari_localview_2023}. We build on that agenda in several ways: by expanding coverage beyond YouTube-based meetings, by recovering speaker-level participation and issue-level structure from raw transcripts, by linking speakers to voter and property records, and by validating each major stage of the extraction pipeline against human-coded labels. In that sense, the paper contributes both new substantive evidence on local participation and a reusable measurement framework for studying local political speech at scale.

\section{Data}
\begin{figure}[ht!]
    \centering
    \includegraphics[width=.9\linewidth]{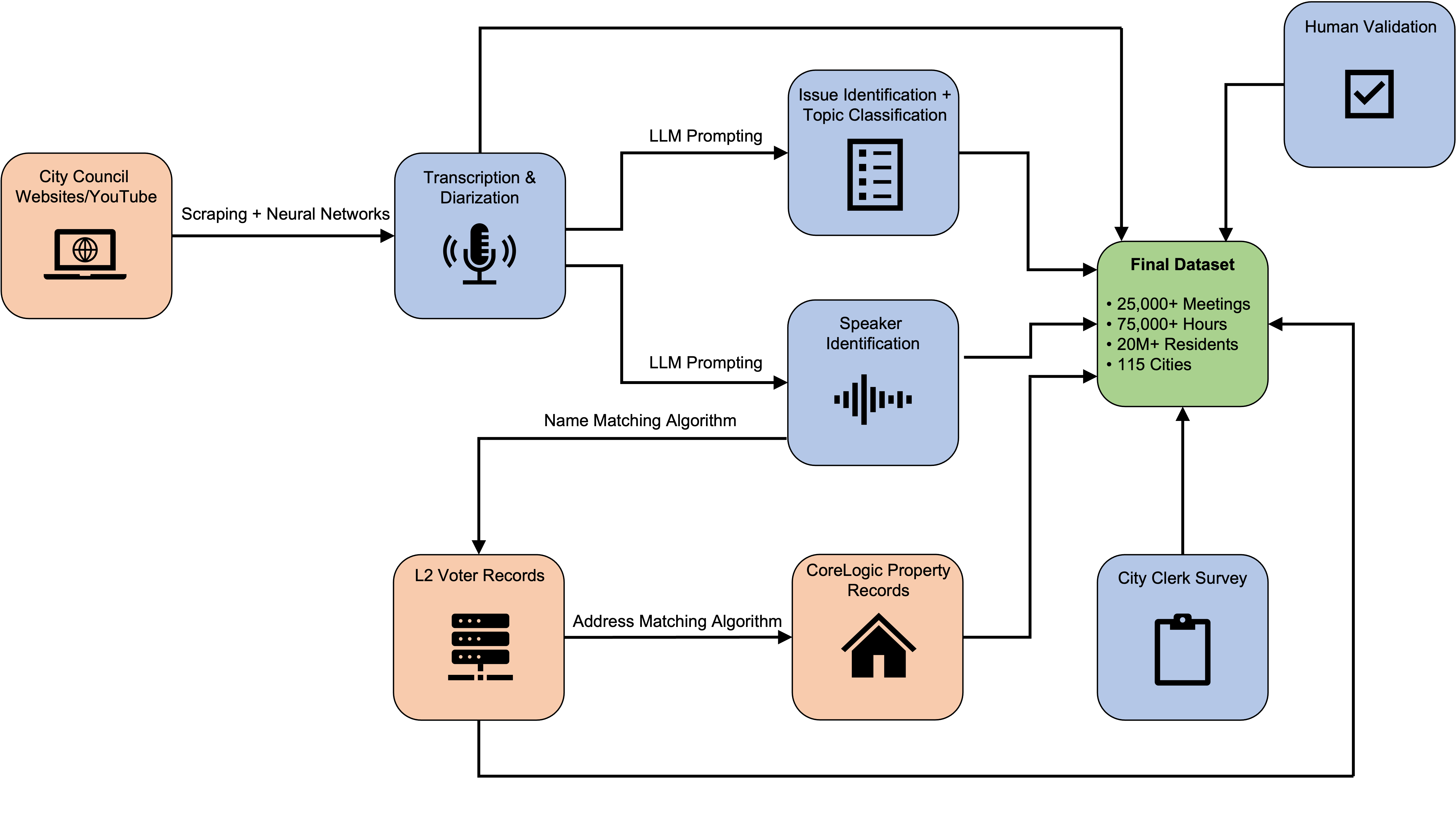}
    \caption{Our data construction pipeline. Orange tiles denote data sources we ingest, while blue tiles denote processed datasets we construct. The final merged dataset is in green.}
    \label{fig:data_pipeline}
\end{figure}

A core contribution of this paper is the construction of a comprehensive, novel dataset of city council meeting transcripts from 115 cities in California. The dataset covers over 25,000 city council meetings between 2015 and 2024, representing more than 75,000 hours of recorded deliberation. These cities range from large metropolitan areas such as Los Angeles and San Francisco\footnote{As San Francisco is both a city and a county, its meetings are technically those of the County Board of Supervisors rather than a city council. They are included regardless.} to mid-sized and smaller jurisdictions like Palo Alto and Los Gatos, together accounting for over 20 million residents or about 52\% of the state’s population. We augment these data by matching our compiled transcripts with L2 voter record data \citep{labels__lists_inc_l2_l2_2025}, CoreLogic property ownership \citep{corelogic_nz_limited_and_cotality_cotality_2024}, and a survey of city clerks we conducted to determine remote meeting availability during and after the COVID-19 pandemic. An overview of our data pipeline can be found in \autoref{fig:data_pipeline}. This section describes each component of this pipeline.

\subsection{Scraping, Transcription, \& Diarization}
We begin by scraping archived meeting audio and video files from municipal websites and official YouTube channels. While towns are required to make recordings of their city council meetings available to the public, there is significant heterogeneity in their choice of channel. Past work has scraped meetings from YouTube (most notably the \textit{LocalView} dataset, compiled by \citet{barari_localview_2023}), but we found that this significantly limits coverage, both within and across towns. By expanding our dataset to include additional channels, we remove potential sources of confounding via selection into certain meeting channels and dramatically expand coverage (see \autoref{tab:channel_coverage} in \autoref{apx:addl} for details). Our resulting dataset covers a wide array of cities, with significant heterogeneity across size (in terms of both geography and population), age, racial characteristics, income, and property values. \autoref{tab:sample_demos} summarizes these characteristics, with the final column displaying statewide values for reference. For a view on the geographic spread of our collected data, see \autoref{fig:map} in \autoref{apx:addl}. Since our selected sample contains all of California's largest cities and misses the long tail of small municipalities in the state, cities in our sample exhibit somewhat higher population densities, percentages of Black residents, incomes, and home values, and correspondingly exhibits somewhat lower percentages of Hispanic residents.

\begin{table}[H]
\caption{Demographics of sampled municipalities}
\label{tab:sample_demos}
\centering
\begin{tabular}{lccc|c}
\toprule
 & {Mean} & {Min} & {Max} & {California} \\
\midrule
Population & 178,736 & 30,884 & 3,878,718 & 39,431,263\\
Population Density & 5,629 & 1,401 & 17,720 & 2,607\\
Median Age & 39 & 25 & 48 & 38.4\\
Percent Over 65 & 12 & 7 & 18 & 12\\
Percent White & 37 & 6 & 78 & 38\\
Percent Hispanic & 38 & 5 & 84 & 41\\
Percent Black & 6 & 0 & 41 & 5\\
Median Income & 114,608 & 54,238 & $>$250,000 & 100,149\\
Median Home Value & 948,879 & 281,200 & $>$2,000,000 & 759,500\\
\bottomrule
\end{tabular}
\begin{minipage}{0.85\textwidth}
\vspace{4pt}\footnotesize 
\textit{Notes:} Values obtained from US Census 2024 American Community Survey 1-Year Estimates. Population densities are expressed in persons per square mile of land area. Income and home value estimates are expressed in USD and are truncated above \$250,000 and \$2,000,000 in the data, respectively. Mean values for these variables reflect this truncation.
\end{minipage}
\end{table}

We then generate transcripts using OpenAI's large-vocabulary speech-recognition model, Whisper \citep{radford_robust_2022}, providing a mapping from timestamps to text. We simultaneously identify distinct speakers in the recorded audio via neural network-based diarization framework, pyannote \citep{bredin_pyannoteaudio_2019}, which detects distinct speaker voiceprints and assigns each detected speaker a unique meeting-level ID. This provides a mapping from timestamps to speaker, which we then merge with the transcript data to provide a speaker-level text record. We then implement filters to remove extremely short ($<$15 minutes) and extremely long ($>$15 hours) meetings that are unlikely to provide informative content.

\subsection{Issue Identification \& Topic Classification}
To further structure our data, we segment meetings into discrete ``issues'': continuous stretches of discussion focused on a specific policy topic or agenda item. We prompt an LLM (Gemini 3 Flash) to extract, for each issue: (1) a short title and summary; (2) whether the issue was agendized (i.e. was included on the published meeting agenda) or was raised in unstructured public comment periods; (3) whether the city council and/or members of the public engaged with it; (4) whether a vote was held on the issue; (5) the vote outcome (e.g., number of votes for vs. against); and (6) the stage at which the vote occurred (e.g. final passage, a motion to table discussion to the next meeting, etc.). See \autoref{apx:llm_prompts} for prompting details. Across all meetings, we identify roughly 320,000 issues (about twelve per meeting). 

We then classify each issue into one of ten thematic categories via LLM discovery \citep{modarressi_causal_2025, liu_llm-guided_2025}. We identify these topics through a two-stage procedure. First, because the full set of issues summaries would exceed a single context window, we randomly shuffle all issues and their summaries and pass them to an LLM into eight context-window sized chunks. For each chunk, we prompt the LLM to inductively identify five to ten recurring city-governance topics, including names, descriptions and representative examples. We then provide the chunk-level topic lists to a second LLM prompt, which consolidates overlapping categories into a fixed taxonomy of ten topics. We then use an LLM to classify issues into each of these ten topics. The resulting ten topics capture the dominant domains of city‐council activity—from budgets and land use to policing and infrastructure—with stable frequencies across cities and years. \autoref{tab:issue_topics} illustrates the learned topics, along with the number of issues classified as belonging to that topic. \autoref{tab:topic_issue_examples} in \autoref{apx:addl} provides examples of common issues within each topic.

\begin{table}[H]
\caption{Breakdown of identified issues of discussion by learned topic classification}
\label{tab:issue_topics}
\centering
\begin{threeparttable}
\begin{tabular}{lcc}
\toprule
{Topic} & {\# Issues} & {\% of Total Issues}\\
\midrule
Governance \& Administration\tnote{a} & 82{,}508 & 27.4\\
Infrastructure & 45{,}659 & 15.2\\
Community Services & 32{,}408 & 10.8\\
Land Use \& Zoning & 28{,}087 & 9.3\\
Housing \& Homelessness & 26{,}453 & 8.8\\
Public Safety & 25{,}525 & 8.5\\
Social Equity & 19{,}039 & 6.3\\
Economic Development & 15{,}961 & 5.3\\
Environmental Sustainability & 14{,}291 & 4.7\\
Public Health & 11{,}446 & 3.8\\
\midrule
Total & 301{,}377 & 100\\
\bottomrule
\end{tabular}
\begin{tablenotes}[flushleft]
\footnotesize
\item[a] The full title of this topic is ``Municipal Governance, Finance, and Labor Relations'' and as such, the topic also contains matters related to city budget-setting.
\end{tablenotes}
\end{threeparttable}
\end{table}

We validate this procedure against hand-labeled ground truth from seven randomly sampled meetings, observing high levels of agreement between human labelers and the LLM on issue agendized status, vote occurrence, vote tallies, and issue-level precision and recall. See \autoref{apx:llm_name_valid} for details.

\subsection{Name Extraction}
To augment our speaker-level text records, we employ the same LLM to infer speaker names and roles from linguistic cues and context contained in the produced transcripts. This task is made feasible by the fact that council meetings are highly formalistic, and almost always require public speakers to identify themselves at the beginning of their speech. We pass a meeting transcript to the LLM and prompt it to return a mapping from speaker ID (output by the diarization process) to the speaker's full name, whether or not they are a member of the government, and whether or not they speak on behalf of a group. See \autoref{apx:llm_prompts} for prompting details. Manual validation against 1,300 hand-labeled speakers shows 87\% agreement on government/member of public status of the speaker and relatively high accuracy in retrieving the names of public speakers in a given meeting. Among speakers where both the LLM and human labelers identified a name, last-name exact match rates reach 74\%, with most disagreements reflecting misattribution across speaker segments rather than fabricated names.\footnote{We focus on last name match accuracy because our fuzzy matching algorithm prioritizes last name similarity.} Fewer than 2 percent of LLM-only names ultimately matched a voter record, suggesting that potentially hallucinated names rarely propagate into downstream analyses. Additional details on the hand-validation exercise can be found in \autoref{apx:llm_name_valid}.

\subsection{Voter Registration \& Property Ownership}
We supplement our speaker-level speech data with demographic information about speakers drawing from two datasets. First, we link identified public speakers to the L2 voter registration data \citep{labels__lists_inc_l2_l2_2025} using a custom fuzzy-matching algorithm which takes into account possible errors in spelling from the transcription process. The algorithm prioritizes exact last-name and locality matches, returning a unique candidate when confidence exceeds a threshold. See \autoref{apx:l2_match_algo} for further details on the matching algorithm. We achieve an overall match rate of approximately 70\%.\footnote{This is a relatively high matching rate given the messy nature of transcripts and the scale of the data. For example, \cite{sahn_public_2025} achieves a match rate of 48\% when matching San Francisco Planning Commission public commenters to the voter file; \cite{donahue_politics_2023} achieves a match rate of 68\% matching police officers to a voter file using name and date of birth; \cite{einstein_who_2019} achieve a match rate of 83\% matching Boston metropolitan area meeting participants to the voting file.} To investigate the possibility of poorer matching on non-white names, we run all participant names through a transformer-based model for predicting race and ethnicity (ethnicolr, \citet{chintalapati_predicting_2023})

In \autoref{tab:l2_match_balance}, we check balance between matched and unmatched names by calculating a standardized difference. We observe reasonable balance across broad origin groups, though our sample size is so large that even small differences are statistically significant. The worst match is among African-origin names (e.g., ``Salah El-Bakri'' or ``Zuzka Ejena''), which compose 3.7\% of the matched sample and 6.2\% of the unmatched sample. Otherwise, the matching algorithm performs quite consistently across names by predicted racial/ethnic origin. Matched voter records provide age, gender, party registration, and race/ethnicity estimates, allowing us to benchmark meeting participants against the registered electorate in each city-year.

\begin{table}[H]
\caption{Balance of predicted speaker name origin across L2 match status (Panel A); balance of registered voter demographics across CoreLogic match status (Panel B)}
\label{tab:l2_match_balance}
\centering
\begin{tabular}{lcccc}
\toprule
\textbf{Variable} & \textbf{Std.\ Diff} & \textbf{SE} & \textbf{N Matched} & \textbf{N Unmatched} \\
\midrule
\multicolumn{5}{l}{\textbf{Panel A: Name to L2 Name Match}}\\
African & -0.1260 & 0.0013 & 4{,}864 & 4{,}022 \\
White non-Hispanic & 0.0787 & 0.0024 & 79{,}160 & 37{,}514 \\
Other & -0.0211 & 0.0001 & 4 & 17 \\
Hispanic & -0.0120 & 0.0016 & 10{,}374 & 5{,}356 \\
Asian & -0.0048 & 0.0017 & 13{,}094 & 6{,}603 \\
\midrule
\textit{Total} & & & 107{,}496 & 53{,}512 \\
\midrule
\multicolumn{5}{l}{\textbf{Panel B: L2 Address to CoreLogic Address Match}} \\
Female & $0.004$ & $0.000$ & 22{,}983{,}709 & 16{,}255{,}611 \\
Democrat & $-0.008$ & $0.000$ & 21{,}280{,}428 & 15{,}315{,}956 \\
White non-Hispanic & $-0.004$ & $0.000$ & 18{,}653{,}470 & 13{,}267{,}440 \\
Age & $-0.030$ & $0.004$ & 44{,}272{,}805 & 31{,}596{,}330 \\
Big City & $-0.120$ & $0.000$ & 14{,}170{,}481 & 11{,}919{,}618 \\
\midrule
\textit{Total} & & & 44{,}299{,}414 & 31{,}625{,}802 \\
\bottomrule
\end{tabular}
\begin{minipage}{0.85\textwidth}
\vspace{4pt}\footnotesize 
\textit{Notes:} An observation is a person-year within the L2 data. Big City is defined as residing in a city with a population $>$ 500,000: i.e., Sacramento, San Francisco, Fresno, San Diego, San Jose, \& Los Angeles. Standardized differences $< 0.1$ indicate acceptable covariate balance \citep{rosenbaum_constructing_1985}.
\end{minipage}
\end{table}

We further match speakers to CoreLogic property-tax assessor files \citep{corelogic_nz_limited_and_cotality_cotality_2024} by the address reflected in their voter registration obtained from L2. This allows us to identify homeownership status of public participants. We are able to match 81\% of matched L2 speakers to CoreLogic addresses. As Panel B of \autoref{tab:l2_match_balance} demonstrates (using L2's demographic data), we are able to achieve an overall match rate of 58\% between the registered voter populations of the 115 towns in our sample (as given by L2) and CoreLogic property record data. Most demographics appear well-balanced, with the exceptions of ``Age" and ``Big City" - our matched sample is biased slightly towards younger residents and residents from outside of the six largest cities in the state.

\subsection{Remote Participation}\label{sec:zoom_data}
Ordinarily, city council meetings take place in person, with residents delivering remarks at a podium in council chambers. The COVID-19 pandemic abruptly disrupted this routine. In March 2020, California implemented statewide emergency orders suspending in-person quorum and posting requirements, allowing city councils to convene and accept public comment remotely for the first time. Within weeks, most cities in our sample adopted some form of remote public participation, such as Zoom or call-in phone lines. In this paper, however, we define \emph{remote public participation} narrowly to mean \emph{live remote participation}: residents must be able to participate synchronously as part of the meeting itself, whether by video or live telephone. Cities that accepted only written emails, e-comments, or prerecorded voicemail submissions are therefore not coded as offering remote public participation. While the onset of live remote access was nearly simultaneous across jurisdictions, the return to in-person-only participation was highly staggered. Some cities reinstated in-person-only public comment by mid-2021, while others continued to allow live remote participation in hybrid or fully remote formats for years.

We systematically document this variation using two complementary sources. First, we conducted an email-based survey of city clerks across all 115 municipalities in our sample, asking when remote participation was introduced, modified, and discontinued for both the public and councilmembers. Second, we manually reviewed hundreds of meeting videos, agendas, and city websites to verify whether each meeting offered a live remote option for public comment. These two sources together yield a fine-grained panel of meeting-level participation modes, categorizing each meeting as ``in-person only", ``hybrid", or ``fully remote" for both the council and the public. 

Summary statistics of our compiled remote public participation data are found in \autoref{tab:zoom_coverage}. Approximately 83\% of towns in our sample offer live remote public participation at some point during the pandemic, of which roughly two thirds have since eliminated that option. On average, towns which eliminated previously offered live remote public participation do so after approximately 30 months, or 2.5 years, yet we observe substantial heterogeneity in exact timing of this removal.

\begin{table}[H]
\caption{Coverage of remote public participation data}
\label{tab:zoom_coverage}
\centering
\small
\begin{tabular}{lc}
\toprule
 & {N} \\
\midrule
\multicolumn{2}{l}{\textbf{Panel A: Sample funnel}} \\
Municipalities in sample & 115 \\
\quad Less: email or voicemail only (no live remote access) & 12 \\
\quad Less: insufficient information to code participation format & 7 \\
Ever offered live remote public participation & 96 (83\%) \\
\midrule
\multicolumn{2}{l}{\textbf{Panel B: Status Among Towns That Ever Offered Live Remote Participation}} \\
Returned to in-person-only public comment & 61 (64\%) \\
Still offering remote public comment (as of survey date) & 35 (36\%) \\
\midrule
\multicolumn{2}{l}{\textbf{Panel C: Timing of return to in-person (among towns that ended remote access)}} \\
Earliest return to in-person & Jan\ 2021 \\
Mean months of remote public access & 30.9 \\
\midrule
\multicolumn{2}{l}{\textbf{By year of return to in-person:}} \\
\quad 2021 & 16 \\
\quad 2022 & 16 \\
\quad 2023 & 21 \\
\quad 2024 & 8 \\
\bottomrule
\end{tabular}
\begin{minipage}{0.85\textwidth}
\vspace{4pt}\footnotesize
\textit{Notes:} Cities classified as ``email or voicemail only'' accepted written or recorded submissions but
did not offer live remote access to council proceedings. Duration is measured from the first remote meeting to the last fully-remote meeting before return to in-person-only public comment.
\end{minipage}
\end{table}

A summary of how our sample size changes throughout each step of this data augmentation process can be found in \autoref{tab:waterfall}. Note that we omit the survey of city clerks since, unlike the other filters, this operates at the town level and not the meeting level; since we learn remote access dates for all cities in our sample, this does not lead to a reduction in sample size. Note that we filter to meetings occurring after Jan 1, 2018 due to data limitations for the L2 voter record, for which we have access to annual data beginning in 2018.

\begin{table}[H]
\caption{Sample size over various augmentation steps}
\label{tab:waterfall}
\centering
\begin{tabular}{lccc}
\toprule
{Data Stage} & {\# Meetings} & {\% of Total Meetings} & {\% of Previous Stage}\\
\midrule
Transcription \& Diarization & 25{,}414 & 100 & --\\
Speaker Name ID & 25{,}383 & 99.9 & 99.9\\
Issue/Vote ID & 25{,}235 & 99.3 & 99.4\\
Date $>$ Jan 1, 2018 & 18{,}385 & 72.3 & 72.9 \\
L2 Match & 16{,}185 & 63.7 & 88.0\\
CoreLogic Match & 15{,}462 & 60.8 & 95.5\\
\bottomrule
\end{tabular}
\end{table}

We proceed to leverage this data to characterize the structure and content of city council meetings, analyze the participant pool, and assess the efficacy of remote public participation options as a policy lever to improve representation in city council meetings.

\section{What is Discussed?}
In this section, we present a descriptive analysis of the structure and content of city council meetings in California. We begin by presenting summary statistics, displayed in \autoref{tab:meeting_descriptives}. Cities hold an average of 24 meetings a year, with meetings lasting approximately 3 hours on average. We observe approximately 30 speakers on average per meeting, with nearly half of all speakers being members of the public. Meetings cover an average of 15 issues, and roughly half of those issues receive public comment.

A particularly striking pattern is what happens once issues reach a formal vote. Roughly 43\% of raised issues include a vote by the council, and virtually all votes pass: only about 1\% fail. More surprising, however, is the degree of consensus: 87\% of votes are unanimous. This is far above the rate observed in Congress: only 10\% of votes in the modern Congress have no recorded nays.\footnote{Using Voteview's historical roll-call database, we calculate that only 6,330 of 111,979 House and Senate roll calls in completed Congresses through the 118th had no recorded nays, or 5.65\%. Even restricting the comparison to the modern Congresses for which official online roll-call records are most complete, the rate remains low: among the 101st through 118th Congresses, 3,504 of 34,041 roll calls, or 10.29\%, had no recorded nays. Voteview's roll-call export excludes quorum calls and vacated votes. See \citet{lewis_voteview_2026}. 

State legislatures provide a more mixed comparison and data is more sparse. \citet{jewell_party_1955}, discussing \citet{keefe_parties_1954}, reports that in the 1951 Pennsylvania legislature, 82\% of Senate roll calls and 70\% of House roll calls were unanimous; he also notes that unanimous votes averaged roughly two-thirds of roll calls in Keefe's study of the 1949 and 1951 Illinois sessions.} 

Because this is an unusually high level of agreement, we audited whether it was an artifact of how votes were extracted. One concern was that consent calendars might mechanically inflate unanimity, since a council often casts a single vote on a batch of consent items while our issue-level data can record each item separately. Carefully ensuring to collapse consent-calendar items to one row only brings the unanimous vote share to 85.5\%. A second concern was that unanimity might be driven mainly by purely procedural motions, such as votes to continue an item rather than approve or reject a substantive policy. We therefore use an LLM to separate final votes from procedural votes. This does not explain away the pattern: final votes are actually \textit{more} likely to be unanimous at 88\% compared to procedural votes which are only unanimous 78\% of the time. Finally, we also consider whether the result is driven by ceremonial business, such as proclamations or certificates of recognition.\footnote{We classify
an item as ceremonial if the concatenated issue, summary, and vote-outcome text matches the following
case-insensitive regular expression: \texttt{\textbackslash bproclamations?\textbackslash b|\textbackslash
bcommendations?\textbackslash b|\textbackslash bcertificates? of (recognition|appreciation)\textbackslash
b|\textbackslash btributes?\textbackslash b|\textbackslash bhonoring\textbackslash b|\textbackslash bin
honor of\textbackslash b|\textbackslash bin memoriam\textbackslash b|\textbackslash badjourn(ed|ment)? in
memory\textbackslash b|\textbackslash b(awareness|heritage|history|appreciation|prevention)
month\textbackslash b|\textbackslash bday of remembrance\textbackslash b}, excluding matches that also
contain \texttt{\textbackslash b(covid|coronavirus|emergency proclamation|local emergency|state of
emergency|disaster|disasters)\textbackslash b}.} Votes on these topics are more likely to be unanimous--95\%--but there are so few that removing these votes leaves the unanimity count at 87\%. Thus, the high rate of unanimity is not merely a consent-calendar artifact or a byproduct of procedural housekeeping. It appears to be a central feature of city council decision-making in our sample.

\begin{table}[H]
\caption{Summary statistics of city council meetings}
\label{tab:meeting_descriptives}
\centering
\begin{tabular}{lrrr}
\toprule
\textbf{Panel A: Counts} & \multicolumn{3}{c}{\textbf{Value}} \\
\midrule
Number of meetings & \multicolumn{3}{c}{25{,}414} \\
Number of towns & \multicolumn{3}{c}{115} \\
\midrule
\textbf{Panel B: Meeting Structure} & \textbf{Median} & \textbf{Mean} & \textbf{Std. Dev.} \\
\midrule
Meetings per town & 209 & 221.0 & 119.7 \\
Meetings per town per year & 24 & 25.6 & 15.6 \\
Meeting length (hours) & 2.7 & 3.0 & 1.8 \\
Speakers per meeting & 29 & 34.2 & 22.2 \\
Government speakers per meeting & 16 & 16.7 & 6.6 \\
Public speakers per meeting & 12 & 17.4 & 18.4 \\
\midrule
\textbf{Panel C: Issues and Voting} & \textbf{Median} & \textbf{Mean} & \textbf{Std. Dev.} \\
\midrule
Issues per meeting & 13 & 13.5 & 6.5 \\
Share with public comment & 0.53 & 0.52 & 0.25 \\
Share with a vote & 0.43 & 0.43 & 0.22 \\
Share of votes unanimous & 1.00 & 0.86 & 0.21 \\
Share of votes that fail & 0.00 & 0.01 & 0.06 \\
\bottomrule
\end{tabular}
\end{table}

In \autoref{tab:topic_shares} we examine the varying content of local government meetings based on topics of discussion. We find that most issues focus on matters related to Governance \& Administration, while Land Use \& Zoning sees a markedly higher incidence of close votes (defined as a vote outcome in which one vote flipping would tie or change the outcome) than other categories.

\begin{table}[H]
\caption{Issue shares and voting characteristics by topic}
\label{tab:topic_shares}
\centering
\begin{tabular}{p{4.5cm}cccc}
\toprule
{Topic} & {\% of Issues} & \multicolumn{3}{c}{{Within-Topic Shares (\%)}} \\
\cmidrule(l){3-5}
 & & {Has Vote} & {Unanimous} & {Close Vote} \\
\midrule
Governance \& Admin & 27.4 & 57.9 & 49.7 & 3.2\\
Infrastructure & 15.2 & 47.2 & 42.5 & 1.4\\
Community Services & 10.8 & 28.7 & 25.5 & 1.2\\
Land Use \& Zoning & 9.3 & 55.1 & 44.7 & 3.8\\
Housing \& Homelessness & 8.8 & 41.1 & 35.2 & 1.8\\
Public Safety & 8.5 & 31.0 & 26.9 & 1.3\\
Social Equity & 6.3 & 21.1 & 17.8 & 1.4\\
Economic Development & 5.3 & 42.9 & 34.5 & 3.2\\
Environmental & 4.7 & 37.2 & 32.5 & 1.5 \\
Public Health & 3.8 & 26.4 & 23.0 & 1.1\\
\bottomrule
\end{tabular}
\end{table}

\begin{figure}[H]
\centering
\begin{subfigure}{.5\textwidth}
  \centering
  \includegraphics[width=\linewidth]{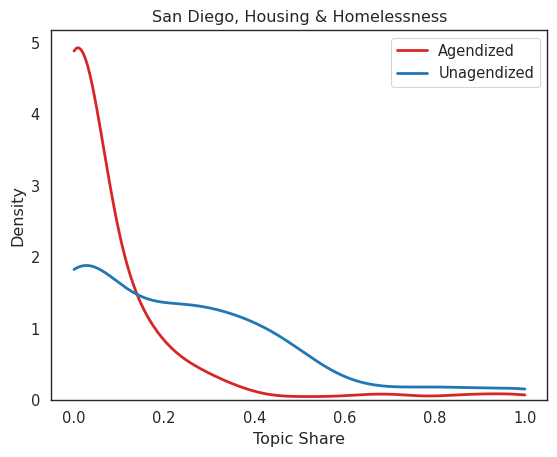}
  \caption{San Diego, Housing \& Homelessness}
  \label{fig:sandiego_homeless}
\end{subfigure}%
\begin{subfigure}{.5\textwidth}
  \centering
  \includegraphics[width=\linewidth]{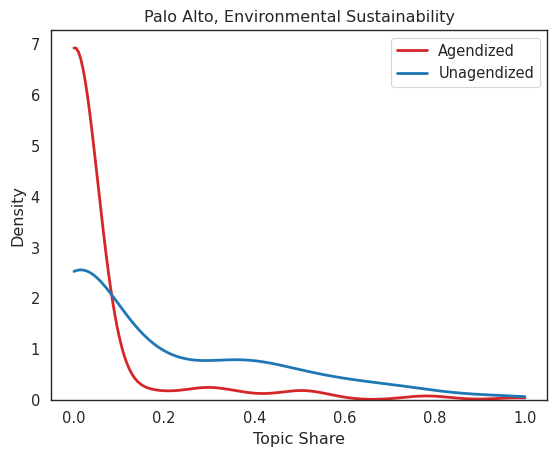}
  \caption{Palo Alto, Environmental Sustainability}
  \label{fig:paloalto_enviro}
\end{subfigure}
\begin{subfigure}{.5\textwidth}
  \centering
  \includegraphics[width=\linewidth]{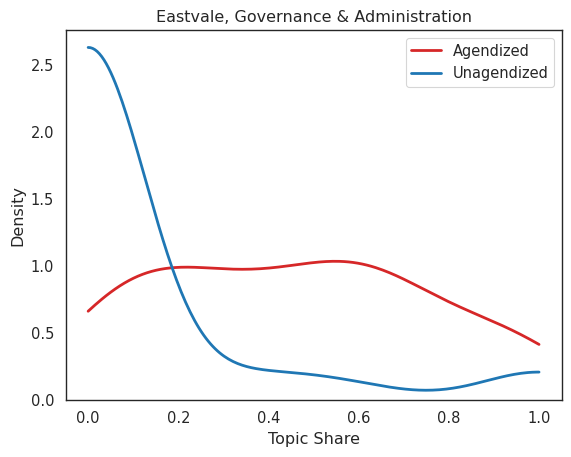}
  \caption{Eastvale, Governance \& Administration}
  \label{fig:eastvale_gov}
\end{subfigure}
\caption{Distribution of agendized discussion (red) and unagendized public comment (blue) topic share for selected topics in selected cities. Smoothed distributions are produced via kernel density estimation.}
\label{fig:topic_towns}
\end{figure}

We can further analyze the variable attention paid to these classes of issues by the city council versus members of the public. We accomplish this by examining the share of speaking time dedicated to a given class of issues within each meeting, separately for issues that are included on the meeting agenda and those that are raised in unstructured public comment sessions. This study reveals notable heterogeneity, with certain cities seeing marked divergence in attention between city councils and members of the public. Two examples are presented in \autoref{fig:topic_towns}. In \autoref{fig:sandiego_homeless} and \autoref{fig:paloalto_enviro}, we see that members of the public place greater emphasis (as measured by speaking time allocation) on Housing \& Homelessness and Environmental Sustainability than the city councils of San Diego and Palo Alto, respectively. By contrast, members of the public in Eastvale under-emphasize discussion of city council operations and administration, as evidenced by \autoref{fig:eastvale_gov}.

\begin{figure}[H]
\centering
\begin{subfigure}{.5\textwidth}
  \centering
  \includegraphics[width=\linewidth]{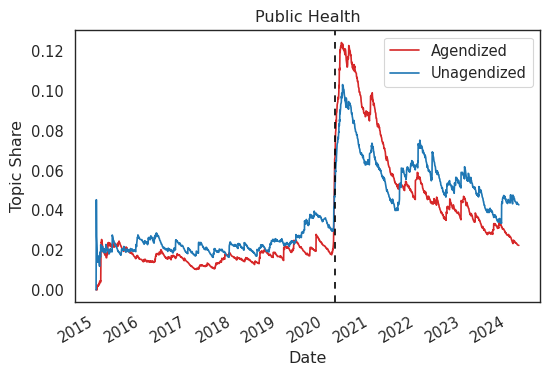}
  \caption{Public Health}
  \label{fig:health_ts}
\end{subfigure}%
\begin{subfigure}{.5\textwidth}
  \centering
  \includegraphics[width=\linewidth]{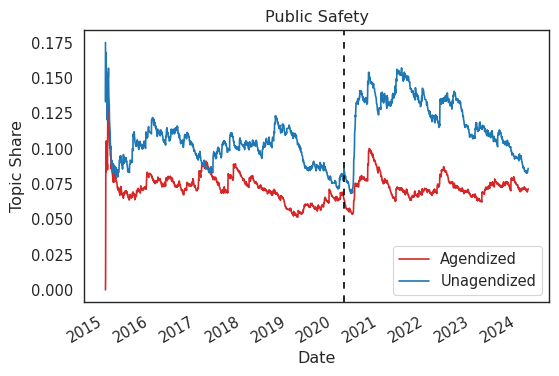}
  \caption{Public Safety}
  \label{fig:safety_ts}
\end{subfigure}
\begin{subfigure}{.5\textwidth}
  \centering
  \includegraphics[width=\linewidth]{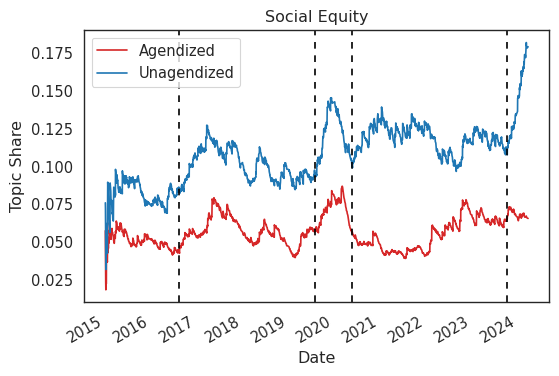}
  \caption{Social Equity}
  \label{fig:equity_ts}
\end{subfigure}
\caption{Exponentially weighted moving averages (with smoothing factor $\alpha=0.01$) of topic shares for selected topics, for agendized discussion (red) and unagendized public comment (blue). Vertical dashed lines in panels (a) and (b) denote the stay-at-home order issued by Gov. Newsom (March 19, 2020). Vertical dashed lines in panel (c) reflect the Milwaukee riots (August 13, 2016), El Paso shooting (August 3, 2019), murder of George Floyd (May 25, 2020), and onset of the Gaza War (October 7, 2023).}

\label{fig:topic_ts}
\end{figure}

Finally, we can examine how the attention paid to various classes evolves over time, separately for agendized issues and unstructured public comment. \autoref{fig:topic_ts} presents these time series for three selected topics. \autoref{fig:health_ts} shows a sharp increase in attention paid to matters related to Public Health with the onset of the COVID-19 pandemic. \autoref{fig:safety_ts} shows a similar sharp increase for issues related to Public Safety, but notably the increase is much larger in magnitude for unagendized public comment than agendized discussion. \autoref{fig:equity_ts} shows a pattern of cyclical increases and decreases in discussion related to Social Equity over time, with clear spikes and regions of elevated focus following major events related to civil rights and racial or religious violence.

\section{Who Participates?}\label{sec:who}
In this section, we characterize the pool of public participants in city council meetings. We find that participants tend to be older, whiter, more liberal, more male, and more likely to be a homeowner than the full population of registered voters. We further find significant heterogeneity in the participation patterns across cities, with increased public participation correlated with city-level renter share, Gini coefficient, level of education, median income, and share of White residents. Cities with larger populations and larger shares of Black residents, meanwhile, see lower rates of public participation. We continue to characterize the subpopulation of repeat participants, individuals who attend multiple city council meetings in the same city. We find that most participants in city council meetings are in fact repeat participants, and that repeat participants are similarly more likely to be older, whiter, more liberal, more male, and less likely to be a homeowner than the participant pool as a whole. Finally, we show that participants select into discussion based on the topics of meeting agendas, with discussion around Land Use \& Zoning, Social Equity, and Housing \& Homelessness being significant drivers of attendance across demographic groups. Collectively, these findings provide a comprehensive overview of who participates in city council meetings in California.

\subsection{The Individual-Level Representation Gap}

A central question in assessing city council meetings as venues for deliberative democracy is whether the individuals who participate resemble the broader public they ostensibly represent. Prior research (e.g., \citet{fowler_what_2025}) has shown that the voting population skews older and whiter relative to the general population. Yet far less is known about the population who takes the additional step beyond voting to actually show up to speak in local government meetings.

The most systematic prior evidence comes from \cite{einstein_who_2019}, who hand-coded participants in planning and zoning board meetings in 97 cities and towns in metropolitan Boston between 2015 and 2017. They successfully matched 83 percent of 3,123 speakers to voter records and found that meeting participants were, on average, eight years older and more likely to be male than the general voting population. In a detailed case study of one town, they further linked 85 participants to property records and found that homeowners comprised 78 percent of attendees, compared to 61 percent of the town’s population.

In addition to adding geographic and temporal diversity, our dataset scales that approach by nearly two orders of magnitude: we match over 100,000 participants across 115 California municipalities to voter registration and property data for the same cities. Approximately 1 in 1,000 registered voters participate in their local city council meeting in a given year. \autoref{tab:balance_simple_clean} reports simple differences in means between participants and non-participants. As in the Boston sample, meeting attendees in our data are older (+4 years), less likely to be female (–5 percentage points), more likely to be Democrats (+6 p.p.), and substantially more likely to be of White non-Hispanic origin (+16 p.p.). Participants are also somewhat more likely to be homeowners (+4 p.p.), although as noted earlier, our matched ownership sample is biased toward smaller towns, providing somewhat high baseline ownership rates.\footnote{For example, the Census Bureau estimates the homeownership rate in California to be around 55\% \citep{bureau_housing_nodate}. However, the Public Policy Institute of California finds the rate to be closer to 73\% among likely voters, suggesting that our overall mean of 70\% may not be a significant overstatement \citep{baldassare_californias_2025}.} 

These gaps are substantively large even relative to familiar forms of political participation. One benchmark is voting itself. Although this comparison is imperfect---our benchmark is the registered voter population, whereas Census benchmarks compare voters to the citizen voting-age population (CVAP)---the racial skew in meeting participation is considerably larger. Our White non-Hispanic gap is 16.4 percentage points, compared with 6.9 percentage points among 2022 voters and 4.2 percentage points among 2020 voters.\footnote{These Census benchmarks compare the White non-Hispanic share of voters to the White non-Hispanic share of the citizen voting-age population (CVAP), rather than to the registered voter population. In the CPS Voting and Registration reports, the corresponding White non-Hispanic overrepresentation is 4.2 percentage points in 2020 and 6.9 percentage points in 2022. See \citet{fabina_voting_2022} and \citet{fabina_voting_2024}.} A second benchmark is campaign giving. Here, if anything, the donor class appears even more selective: in linked donor-voter data, Democratic donors are 19.0 percentage points whiter than non-donors \citep{hill_representativeness_2017}. In this sense, city council speakers appear less representative than the electorate and closer to the donor class than to the voting public.\footnote{For scale, about 66.8\% of adult citizens voted in 2020, whereas roughly 8.5\% appeared as federal donors in the 2020 cycle. Because these are national benchmarks and the donor figure is measured over a presidential cycle, we treat them as rough reference points rather than directly comparable participation rates.}

\begin{table}[H]
\caption{Raw difference in means between participants and non-participant registered voters}
\label{tab:balance_simple_clean}
\centering
\begin{tabular}{lcccccc}
\toprule
{Variable} & \multicolumn{2}{c}{{Participants}} & \multicolumn{2}{c}{{Non-Participants}} & {Difference} \\
\cmidrule(lr){2-3} \cmidrule(lr){4-5}
 & {N} & {Mean} & {N} & {Mean} &  \\
\midrule
Age & 107{,}788 & 52.596 & 75{,}761{,}347 & 48.133 & 4.463$^{***}$ \\
Female & 107{,}077 & 0.478 & 74{,}384{,}571 & 0.526 & -0.049$^{***}$ \\
Democrat & 107{,}788 & 0.537 & 75{,}761{,}347 & 0.482 & 0.055$^{***}$ \\
White non-Hispanic & 99{,}214 & 0.623 & 69{,}394{,}181 & 0.459 & 0.164$^{***}$ \\
Homeowner & 69{,}320 & 0.740 & 44{,}230{,}094 & 0.703 & 0.037$^{***}$ \\
\bottomrule
\end{tabular}
\vspace{0.5em}
\end{table}

To account for local composition and year effects, \autoref{tab:part_logit} reports results from an individual-year-level logistic regression with city and year fixed effects. Each coefficient represents the marginal association between a demographic trait and the log-odds of that person participating in a council meeting. The results confirm the descriptive patterns: older, Democratic, and White non-Hispanic-origin residents are significantly more likely to participate, while women remain less likely to do so even after controlling for location and time. Homeownership is also positively associated with participation, though its effect is smaller in magnitude once other demographics are held constant. Together, these estimates suggest that the demographic biases in participation are systematic and persistent across cities and years.

\begin{table}[H]
\caption{Logistic regression of participation on demographic characteristics}
\label{tab:part_logit}
\centering
\begin{tabular}{lcc}
\toprule
 & (1) & (2) \\
\midrule
Age                  & $0.009^{***}$ & $0.008^{***}$\\
                     & $(0.001)$ & $(0.001)$\\
Democrat             & $0.241^{***}$ & $0.246^{***}$\\
                     & $(0.018)$ & $(0.018)$\\
White non-Hispanic   & $0.545^{***}$ & $0.566^{***}$\\
                     & $(0.029)$ & $(0.031)$\\
Female               & $-0.211^{***}$ & $-0.216^{***}$\\
                     & $(0.018)$ & $(0.017)$\\
Homeowner            & -- & $0.130^{***}$\\
                     & -- & $(0.025)$\\
\midrule
N (observations)     & 68{,}257{,}730 & 40{,}063{,}858\\
N (cities)           & 115 & 115 \\
$\bar{y}$ (mean participation) & 0.0014 & 0.0016\\
City fixed effects   & Yes & Yes \\
Year fixed effects   & Yes & Yes \\
\bottomrule
\end{tabular}
\begin{minipage}{.85\textwidth}
\vspace{4pt}\footnotesize
\textit{Notes:} Coefficient estimates from a logistic regression of meeting participation on demographic characteristics with city and year fixed effects. Standard errors clustered at the city level in parentheses.  Significance levels: $^{*}p<0.10$, $^{**}p<0.05$, $^{***}p<0.01$.
\end{minipage}
\end{table}

\begin{figure}[H]
    \centering
    \includegraphics[width=0.75\linewidth]{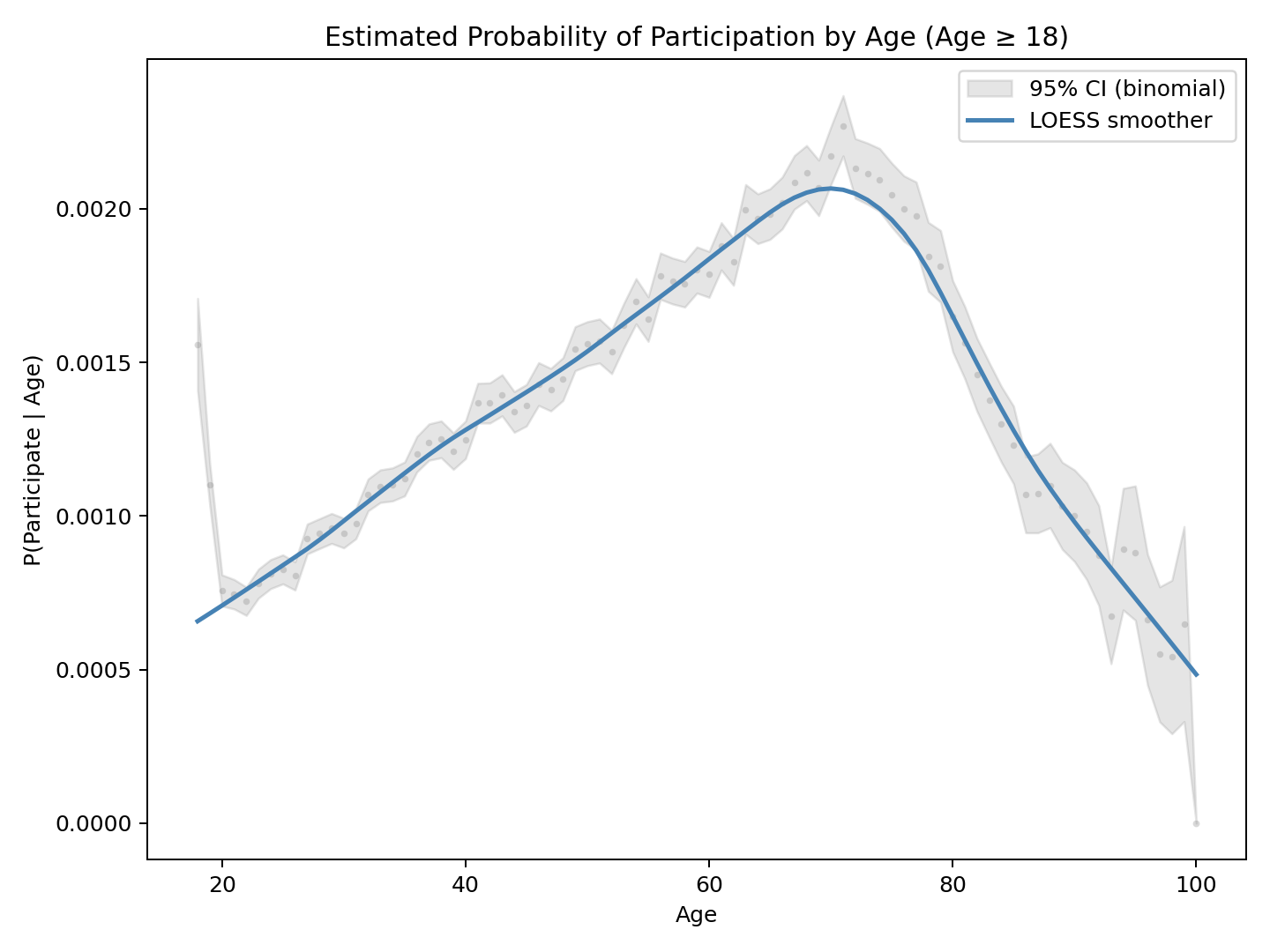}
    \caption{Average participation rates by age}
    \label{fig:prob_particip_by_age}
\end{figure}

Another way to visualize this, at least for age, is to estimate the probability of participation by age among registered voters. As \autoref{fig:prob_particip_by_age} shows, those around 65-75 years old are the most likely to participate in meetings, with a noticeable bump for 18-year-olds, which is likely driven by the frequency of high school students appearing at meetings and the relatively low denominator of registered voters aged 18.

\subsection{City Characteristics and Participation}

\begin{figure}[H]
    \centering
    \includegraphics[width=0.95\linewidth]{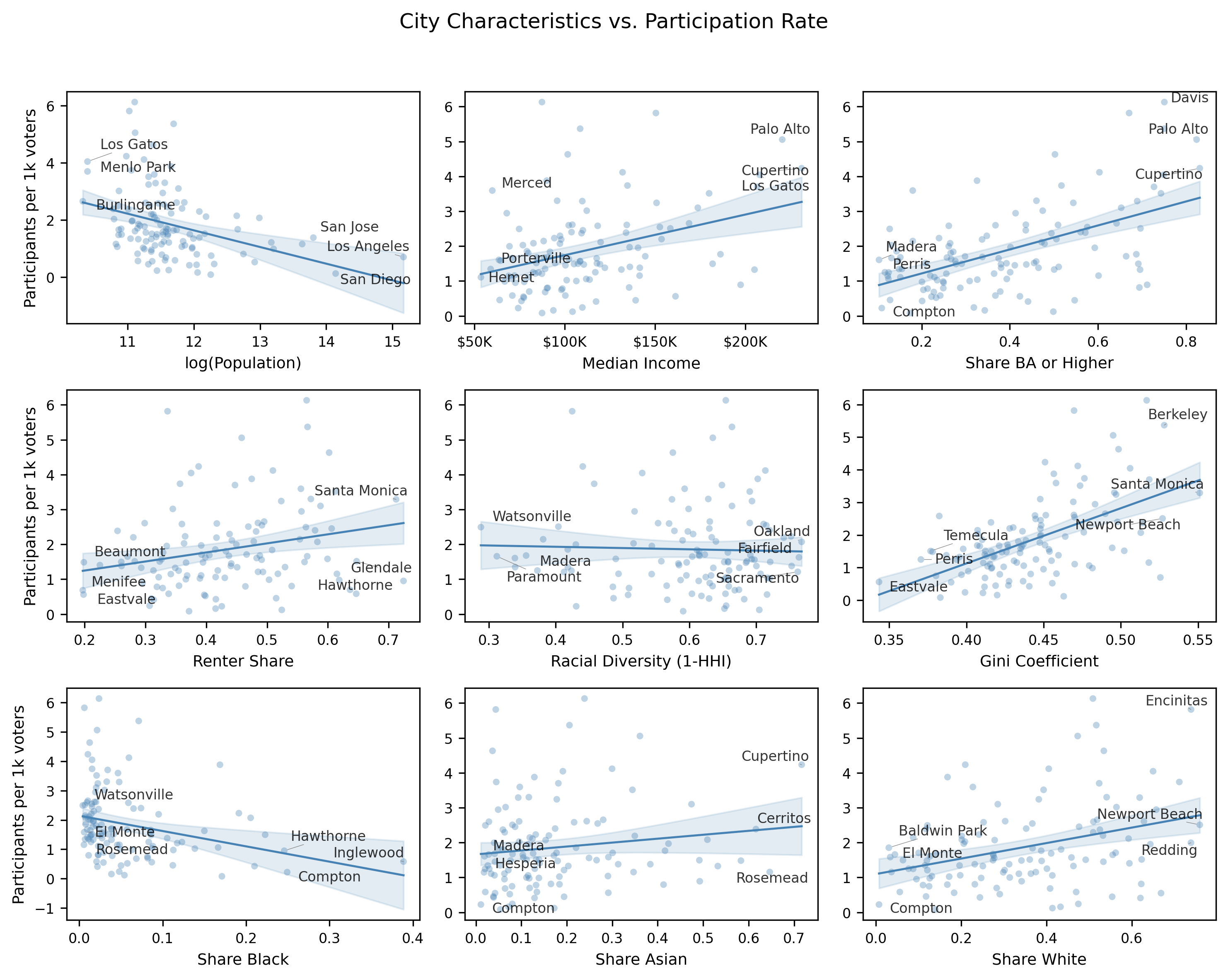}
    \caption{City characteristics and per-capita participation rates}
    \label{fig:acs_predictors}
    \begin{minipage}{.85\textwidth}
    \vspace{4pt}\footnotesize
    \textit{Notes:} Each panel plots per-capita participation (matched speakers per 1,000 registered voters) against a city-level characteristic drawn from the American Community Survey. Lines are OLS fits with 95\% confidence bands.
    \end{minipage}
\end{figure}

Having documented extensive individual-level differences between participants and non-participants, we now identify city-level characteristics that are predictive of these differences. \autoref{fig:acs_predictors} plots per-capita participation against six Census-derived covariates. The share of adults holding a bachelor's degree or higher is the strongest single predictor: a one-standard-deviation increase in educational attainment is associated with roughly a doubling of per-capita participation. Median household income and the Gini coefficient show similar positive associations, consistent with participation being concentrated in affluent communities with high human capital. City population exhibits a strong negative relationship---larger cities have substantially lower per-capita rates---which likely reflects both the mechanical denominator effect and the greater anonymity and lower perceived influence of any individual speaker in a large jurisdiction. Racial diversity (measured as $1 - \text{HHI}$) is negatively associated with participation.

Importantly, the demographic biases documented in \autoref{tab:balance_simple_clean} are not artifacts of a few outlier cities. Across nearly every city in our sample, the age gap between participants and non-participants is positive, the White non-Hispanic-origin gap is positive, and the female gap is negative, as reflected on the vertical axes of \autoref{fig:gap_vs_rate} below. The universality of these patterns suggests that the forces generating unrepresentative participation---whether differential costs of attendance, differential stakes, or differential civic norms---operate similarly across very different local contexts.

A natural question is whether cities that attract more participation also attract more \textit{representative} participation. \autoref{fig:gap_vs_rate} plots each city's representation gap against its per-capita participation rate. For age, the relationship is positive: cities with higher participation tend to have \textit{larger}, not smaller, age gaps between participants and non-participants. The White non-Hispanic-origin gap is roughly flat across participation levels, and the female gap, on the other hand, improves in high-participation cities. These patterns suggest that simply increasing the volume of participation may not, on its own, correct representational distortions.

\begin{figure}[H]
    \centering
    \includegraphics[width=0.95\linewidth]{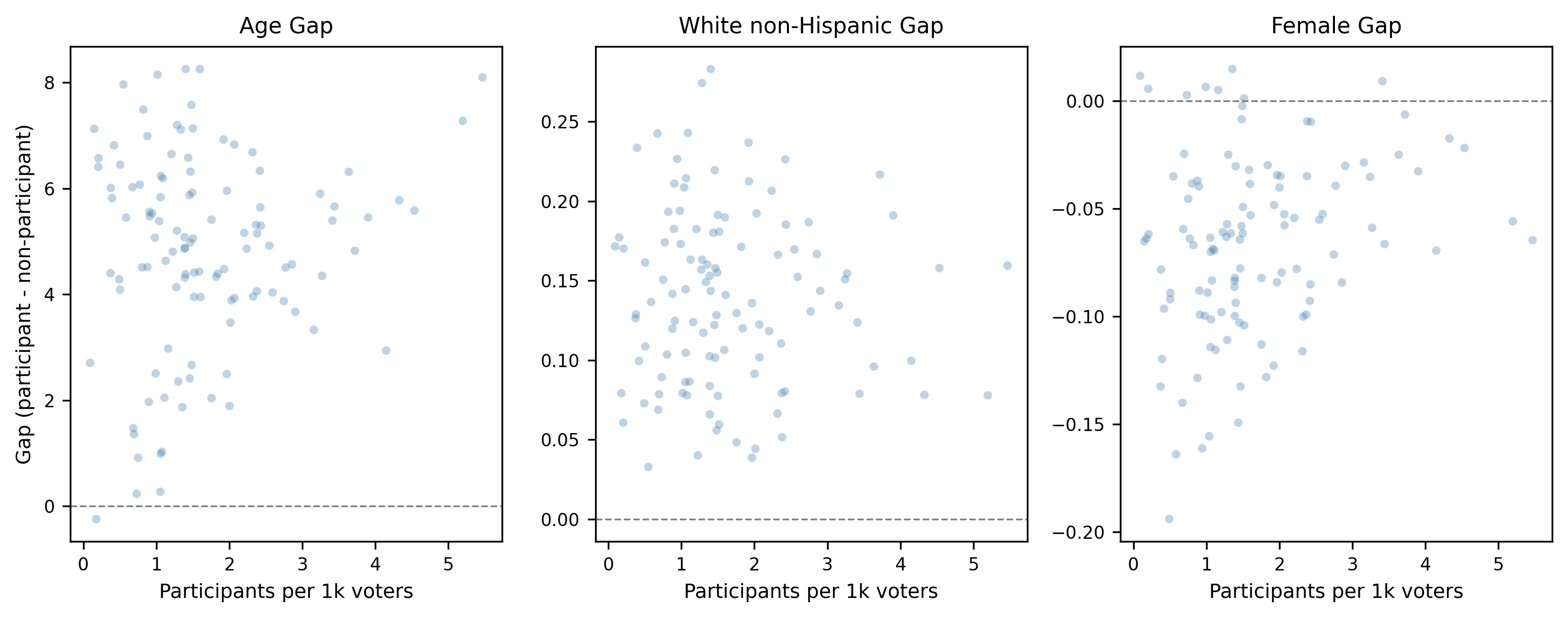}
    \caption{Representation gaps and per-capita participation rates}
    \label{fig:gap_vs_rate}
    \begin{minipage}{.85\textwidth}
    \vspace{4pt}\footnotesize
    \textit{Notes:} Each point is a city. The $x$-axis measures per-capita participation (matched speakers per 1,000 registered voters). The $y$-axis measures the gap between participant and non-participant means for each demographic characteristic.
    \end{minipage}
\end{figure}

\subsection{Repeat Participants}
Among those speakers for whom we are able to obtain a match to the L2 voter record, we can further analyze the frequency with which a given speaker appears repeatedly at meetings within a given town. We do so by classifying participants into one of two categories: ``one-timers" (individuals who participate in exactly one meeting in a given city) and ``repeaters" (individuals who participate in more than one meeting in a given city). We begin by presenting summary statistics of the repeater status of the participant pool. \autoref{tab:repeat_stats} demonstrates that while a significant majority of participants are one-timers (Panel A), most participation instances are driven by repeaters (Panel B), with the median repeater participating in six separate meetings and a long tail of participants appearing many more times than that (Panel C). These facts collectively demonstrate that most individuals who participate in a meeting participate only once, but most participation is perpetuated by a relatively small minority of repeat participants.

\begin{table}[H]
\caption{Summary statistics of repeat participation}
\label{tab:repeat_stats}
\centering
\begin{tabular}{lrr}
\toprule
\textbf{Panel A: Unique Individuals by Status} & \textbf{Value} & \textbf{Percentage}\\
\midrule
One-Timer & 63{,}067 & 73.5 \\
Repeater & 22{,}772 & 26.5 \\
\midrule
\textbf{Panel B: Meeting Appearances by Status} & \textbf{Value} & \textbf{Percentage}\\
\midrule
One-Timer & 63{,}067 & 37.6 \\
Repeater & 104{,}531 & 62.4 \\
\midrule
\textbf{Panel C: Repeater Appearance Counts} & \multicolumn{2}{c}{\textbf{Value}} \\
\midrule
Median & \multicolumn{2}{c}{6.0}\\
Mean & \multicolumn{2}{c}{16.9} \\
Std. Dev. & \multicolumn{2}{c}{34.2}\\
\bottomrule
\end{tabular}
\end{table}

\begin{figure}[H]
    \centering
    \includegraphics[width=0.6\linewidth]{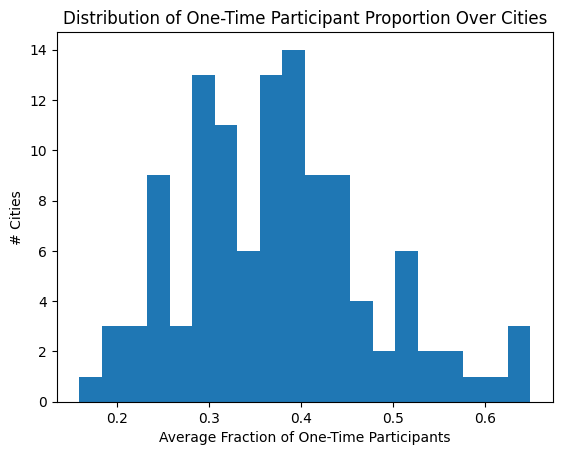}
    \caption{Distribution of first-time participation rates across cities}
    \label{fig:repeat_dist}
\end{figure}

Additionally, we observe significant heterogeneity in the proportion of participants who are one-timers across cities. \autoref{fig:repeat_dist} shows the distribution of this proportion, averaged across all meetings in a given city and demonstrates that these cities behave very differently in their norms regarding repeat participation, with towns ranging from one-time participant rates of 16\% (Baldwin Park) to 65\% (San Diego). 

We can further characterize the difference between repeaters and one-timers by regressing whether a participant is a repeater (i.e. they have either participated previously or will appear again in the future) on their observed demographics, via logistic regression. The results of this regression are shown in \autoref{tab:repeat_logit}.

\begin{table}[H]
\caption{Logistic regression of repeat participation on demographic characteristics}
\label{tab:repeat_logit}
\centering
\begin{tabular}{lcc}
\toprule
 & (1) & (2) \\
\midrule
Age                  & $0.025^{***}$ & $0.012^{***}$\\
                     & $(0.001)$ & $(0.001)$\\
Democrat             & $0.269^{***}$ & $0.279^{***}$\\
                     & $(0.031)$ & $(0.043)$\\
White non-Hispanic   & $0.372^{***}$ & $0.401^{***}$\\
                     & $(0.037)$ & $(0.040)$\\
Female               & $-0.086^{**}$ & $-0.081^{**}$\\
                     & $(0.034)$ & $(0.035)$\\
Homeowner            & -- & $-0.049$\\
                     & -- & $(0.045)$\\
\midrule
N (observations)     & 150{,}192 & 97{,}781\\
N (cities)           & 115 & 115 \\
$\bar{y}$ (mean ``repeater" status) & 0.62 & 0.63\\
City fixed effects   & Yes & Yes \\
Year fixed effects   & Yes & Yes \\
\bottomrule
\end{tabular}
\begin{minipage}{.85\textwidth}
\vspace{4pt}\footnotesize
\textit{Notes:} Coefficient estimates from a logistic regression of repeat participation on demographic characteristics with city and year fixed effects. Standard errors clustered at the city level in parentheses.  Significance levels: $^{*}p<0.10$, $^{**}p<0.05$, $^{***}p<0.01$.
\end{minipage}
\end{table}

These coefficients suggest that repeat participants are more likely to be older, more liberal, have White non-Hispanic ancestry, and be male, findings which are qualitatively similar to those presented in \autoref{tab:part_logit}, suggesting that selection on observables into repeat attendance, conditional on attending once, behaves similarly to selection into attendance in the first place. A key difference, however, is that conditional on attending at least once, homeowner status is (weakly) associated with a lower likelihood of repeat attendance ($p=0.25$). This is consistent with a paradigm in which homeowners are more motivated to attend to speak on an issue directly relating to their property ownership, while non-homeowners who attend meetings do so out of concerns for the broader community and so are more likely to attend repeatedly. However, this effect is not statistically significant and so may be the result of random noise, in which case the selection patterns into repeat participation are identical to those into participation in the first place.

\subsection{Agendas and Participation}
We can additionally study how varying concentrations of topics in meeting agendas result in different participant pools. That is, by merging our L2-matched speakers to our extracted issue identification and topic classifications, we can determine how meetings with various emphasis on different topics draw different types of attendees. In order to reconstruct a proxy for the meeting agenda (which potential participants may use when choosing whether or not to participate) we calculate the percentage of issues in a given meeting belonging to each topic class, among those issues which were determined to be agendized via our LLM-based issue identification task. We then merge these proportions with participant demographics for the given meeting and regress each of our demographics variables on these topic proportions. The resulting coefficient estimates can be found in \autoref{tab:demo_topic_reg}. Notably, we observe that dicsussions around Land Use \& Zoning and Social Equity see high attendance; meetings with greater emphasis on Social Equity and Economic Development see younger participants; and meetings with greater emphasis on Community Services, Social Equity, and Public Health see more female participants.

\begin{table}[H]
\caption{TWFE regressions of participant demographics on agendized topic proportions}
\label{tab:demo_topic_reg}
\centering
\begin{tabular}{lcccccc}
\toprule
 & \# Participants & Age & Democrat & White & Female & Homeowner \\
\midrule
Gov. \& Admin            & $-1.44^{**}$ & $-0.01$       & $-0.01$     & $-0.03^{**}$  & $0.01$       & $-0.03^{**}$\\
Community Services       & $0.50$       & $-0.08$       & $-0.01$     & $-0.01$       & $0.06^{***}$ & $-0.00$\\
Land Use \& Zoning       & $5.31^{***}$ & $0.696$        & $-0.02$     & $0.01$        & $0.03$       & $-0.01$\\
Housing \& Homelessness  & $1.65^{*}$   & $-0.057$       & $0.014$      & $-0.04^*$     & $0.04^{*}$   & $-0.07^{***}$\\
Public Safety            & $-0.76$      & $-0.817$       & $0.002$      & $0.00$        & $0.02$       & $-0.05^{**}$\\
Social Equity            & $4.57^{***}$ & $-1.91^{**}$  & $0.01$      & $-0.10^{***}$ & $0.08^{***}$ & $-0.05^{**}$\\
Econ Development         & $1.45$       & $-3.51^{***}$ & $-0.03$     & $-0.08^{***}$ & $0.01$       & $-0.09^{***}$\\
Environment              & $1.02$       & $0.647$        & $0.051^{**}$ & $0.04$        & $0.02$       & $0.01$\\
Public Health            & $-1.02$      & $-1.75$       & $-0.03$     & $-0.07^{**}$  & $0.11^{***}$ & $-0.02$\\
\midrule
N (observations)     & 16{,}074 & 16{,}074 & 16{,}074 & 15{,}979 & 16{,}066 & 15{,}391\\
N (cities)           & 115 & 115 & 115 & 115 & 115 & 115 \\
City fixed effects   & Yes & Yes & Yes & Yes & Yes & Yes \\
Year fixed effects   & Yes & Yes & Yes & Yes & Yes & Yes \\
\bottomrule
\end{tabular}
\begin{minipage}{.85\textwidth}
\vspace{4pt}\footnotesize
\textit{Notes:} Coefficient estimates from a two-way fixed effected regression of participant demographics on agendized topic proportions with city and year fixed effects. The topic ``Infrastructure" is dropped as the reference topic, to prevent multicolinearity; it was selected as the reference topic as it is the topic with the highest percentage of unanimous votes.  Standard errors are clustered at the city level. Significance levels: $^{*}p<0.10$, $^{**}p<0.05$, $^{***}p<0.01$.
\end{minipage}
\end{table}

\subsubsection{Case Study on Land Use Sentiment and Rhetoric}

One advantage of full transcript data is that it allows us to move beyond asking which issues draw public participation and instead examine how residents frame those disputes once they arrive. We illustrate that added leverage in the domain that generates the highest levels of public engagement in our data: land use and zoning. Across our full sample, 96,604 individual speaker-issue comments—spanning 7,502 distinct agenda items in 6,335 meetings across 115 California cities—were made on issues classified as ``Land Use, Zoning, and Urban Development.'' To study how supporters and opponents of development differ in their expressed views and rhetorical framing, we construct a topic-specific stance measure: for each comment, we provide an LLM with both the issue description and the speaker's text and ask whether the speaker supports or opposes a more development-oriented outcome on a continuous scale from --1 (clearly opposed) to 1 (clearly supportive).

We aggregate this stance measure at the city level to characterize the public mood toward development among a city's most politically active residents. On average, public commenters lean modestly pro-development (mean score = 0.14),\footnote{This estimate likely overstates pro-development sentiment among ordinary residents because our speaker classification distinguishes only between city officials and non-city government speakers. As a result, developers, applicants, and their representatives are included in the ``public speaker'' category rather than separated out from other members of the public.} though with substantial variation (standard deviation = 0.61). Pro-development sentiment was relatively stable from 2015 through 2021, hovering around 0.17--0.19, before declining modestly in 2022 and 2023. This slight softening in expressed pro-development sentiment in the most recent years of our sample may reflect the intensification of community opposition to state-mandated housing policies, though interpretation requires caution given the changing composition of issues that come before councils over time.

To understand how supporters and opponents differ in their rhetoric, we examine which topics are most frequently invoked by each side (\autoref{tab:diff_per1k}) through keyword frequency.\footnote{Keyword frequencies were computed by searching for regular‐expression matches in the transcript text. For instance, we counted mentions of ``parking'' (\texttt{\textbackslash bparking?\textbackslash b}); ``traffic'' (any of \texttt{traffic|congestion|cars?|gridlock}); ``affordability'' (\texttt{affordable|affordability|income}); ``crime'' (\texttt{crime|safety|police|violence|theft|assault|criminal}); ``schools'' (\texttt{school|student|classroom|enrollment}); ``infrastructure'' (\texttt{infrastructure|sewer|water|drainage|utility|utilities|pipes?|electric|power}); ``aesthetics'' (\texttt{aesthetic|appearance|beauty|character|historic|preserv|style|design|architecture|shadow|tall}); and ``environment'' (\texttt{environment|wildlife|animal|bird|tree|pollution|toxic|habitat|greenhouse|climate|hazard}). All expressions are case‐insensitive and bounded by word boundaries (\texttt{\textbackslash b}).} Pro-development speakers are defined as those with a stance score of at least 0.7 (24,516 comments); anti-development speakers are those with a score of --0.7 or lower (13,145 comments).

\begin{table}[H]
\caption{Differences in mentions per 1,000 words by topic}
\label{tab:diff_per1k}
\centering
\begin{tabular}{lccc}
\toprule
{Topic} & {Pro Mean} & {Anti Mean} & {Difference per 1000 Words} \\
\midrule
Affordability & 0.965 & 0.761 & 0.203$^{***}$ \\
Parking & 1.051 & 1.040 & 0.012 \\
Infrastructure & 0.618 & 0.688 & -0.070$^{**}$ \\
Schools & 0.557 & 0.674 & -0.117$^{***}$ \\
Aesthetics & 1.113 & 1.369 & -0.255$^{***}$ \\
Environment & 0.432 & 0.742 & -0.310$^{***}$ \\
Crime & 0.314 & 0.703 & -0.389$^{***}$ \\
Traffic & 0.874 & 1.787 & -0.913$^{***}$ \\
\midrule
Total comments & 24,516 & 13,145 \\
\bottomrule
\end{tabular}
\begin{minipage}{.85\textwidth}
\vspace{4pt}\footnotesize
\textit{Notes:} Stars indicate significance levels based on two-sided Welch $t$-tests. 
$^{*}p<0.1$, $^{**}p<0.05$, $^{***}p<0.01$.
\end{minipage}
\end{table}

The results confirm a clear rhetorical divide. Pro-development speakers disproportionately reference affordability and economic opportunity, while opponents focus on traffic, crime, schools, environmental impacts, and building aesthetics. These word-level differences suggest that each side situates development debates in distinct moral and practical frames---economic necessity and housing need versus neighborhood preservation and quality-of-life concerns.

\section{How Do Access Costs Impact Participation?}

Having observed substantial representation gaps in \autoref{sec:who}, we now turn to a key policy lever available to local governments to improve representativeness of public participation: meeting access costs. We introduce a simple framework for meeting participation in which access costs enter into the individual's participation decision. We then leverage the introduction and subsequent staggered removal of remote public participation options during and after the COVID-19 pandemic as a natural experiment to estimate causal effects. We find that the elimination of remote participation options, corresponding to an increase in participation costs, results in fewer participants per meeting.

\subsection{A Simple Framework for Meeting Participation}\label{sec:framework}

Why does anyone show up to a city council meeting? Adapting the canonical calculus-of-voting framework presented in \citet{riker_theory_1968}, suppose citizen $i$ attends a meeting when $B_i + D_i > C_i$, where $B_i$ captures the expected policy benefit of participation--likely a function of how much is at stake for \textit{i} on the agenda and how likely \textit{i}'s comment is to influence the outcome, $D_i$ captures expressive returns from showing up and feeling heard (or, in the authors' conception, ``civic duty"), and $C_i$ is the cost of attendance: travel time, opportunity cost of an evening, childcare, mobility constraints and so on. Note that $C_i$ likely varies systematically with demographics: older residents with health limitations may face physical costs of attending in person while working-age residents with young children or inflexible jobs may face high opportunity costs concentrated in the evening hours when meetings occur. Homeowners with property values directly at stake may have higher $B_i$ on land use items--but renters with the same $C_i$ may have weaker perceived $B_i$ if they believe the council is less responsive to their preferences \citep{einstein_who_2019, fischel_homevoter_2009}.

Under this framework, a municipal government faced with unrepresentative public participation has two primary policy levers at its disposal: it can alter the topical content of meetings (impacting $B_i$) or it can alter meeting access costs (impacting $C_i$). City councils, however, are largely constrained in what topics of discussion they must cover in a meeting, with procedural matters, budget and finance, and discussion of anodyne issues of governance dominating meetings, as shown in \autoref{tab:issue_topics}. Consequently, the chief discretionary policy lever available to local governments is control over meeting access costs $C_i$, which they can reduce/increase via introduction/removal of a remote participation option. Given the aforementioned heterogeneity in $C_i$ across demographics, the effect on such a change on the composition of the participant pool will depend on who the marginal participant is. 

\subsection{Descriptive Evidence: The Pandemic Shock}

As described in \autoref{sec:zoom_data}, California's emergency orders moved nearly all city council meetings online in March 2020. Of the 101 cities with usable participation data, 96 adopted live remote public comment during the pandemic. The return to in-person formats was highly staggered: 61 of those 96 eventually eliminated remote access — at dates ranging from early 2021 through late 2024 — while 35 continued to offer it through the end of our sample (\autoref{tab:zoom_coverage}). The remaining 19 cities never offered live remote access at all, accepting only email or voicemail submissions, or permitting in-person attendance.

\begin{figure}[H]
    \centering
    \includegraphics[width=0.85\linewidth]{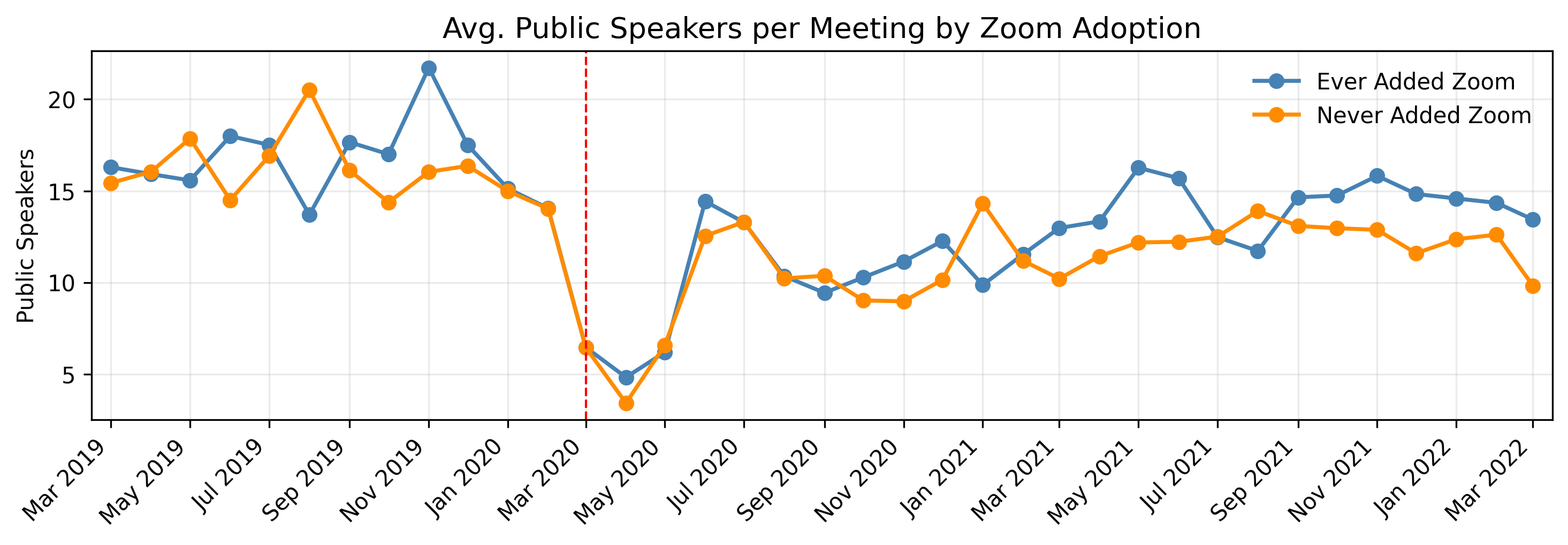}
    \caption{Average public speakers per meeting by remote access adoption}
    \label{fig:final_speaker_count_zoom_on}
\end{figure}

\autoref{fig:final_speaker_count_zoom_on} plots average public speakers per meeting separately for cities that ever adopted remote public access and those that did not.\footnote{The ``Never Added Zoom'' group consists of the 19 cities in our sample that accepted only email or voicemail submissions and never offered live remote participation. These cities serve as a useful visual benchmark but are excluded from the DiD analysis, which relies on within-Zoom-adopter variation in the timing of shutoffs.} Both groups track closely in the pre-period — averaging roughly 15–18 speakers per meeting through early 2020 — and both collapse to around 5–7 speakers in March 2020. The recovery, however, diverges slightly. By mid-2020, cities with remote public participation rebounded to approximately 14 speakers per meeting and sustained that level through 2022. Cities without live remote access recovered slightly more slowly.

\begin{figure}[H]
    \centering
    \includegraphics[width=0.85\linewidth]{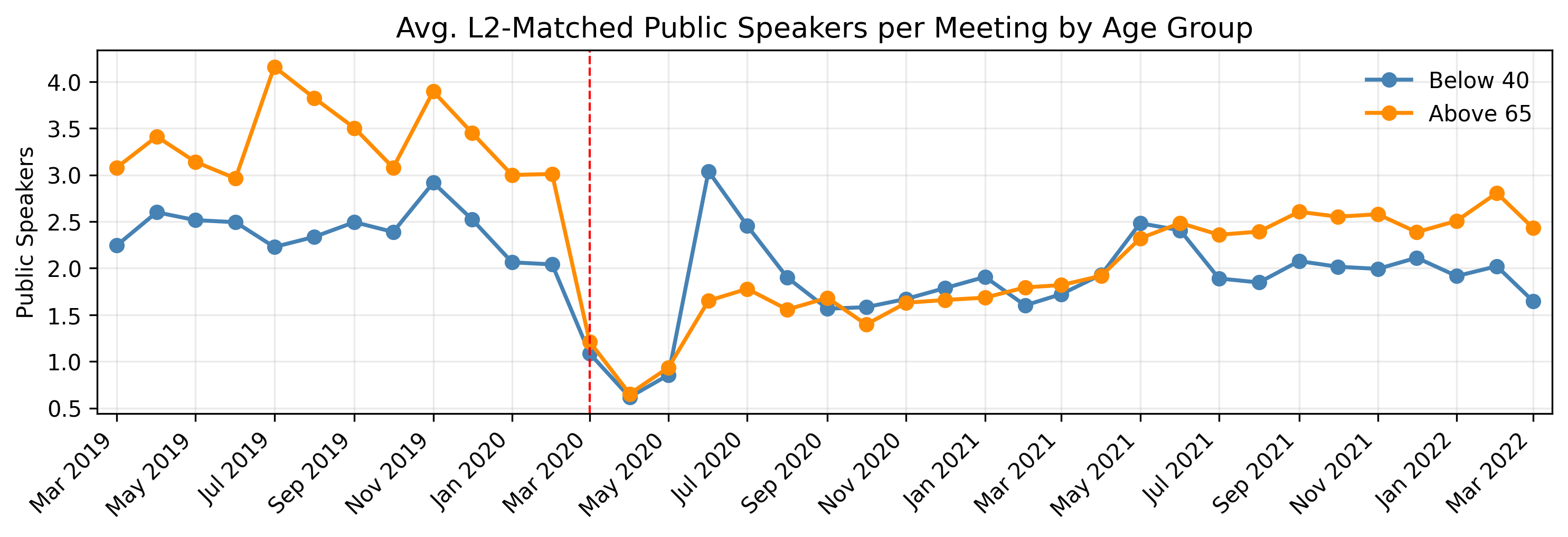}
    \caption{Average public speakers per meeting by age group}
    \label{fig:final_speaker_age_count}
\end{figure}

\autoref{fig:final_speaker_age_count} decomposes the participation time series by age group, plotting average voter record-matched speakers per meeting for residents below 40 and above 65. Before the pandemic, the over-65 group contributed substantially more speakers per meeting — roughly 3.0–4.0 compared to 2.0–2.5 for the under-40 group, a gap consistent with the age skew documented in \autoref{sec:who}. Both groups collapse sharply in March 2020. But the recovery paths differ in a revealing way. The under-40 group recovers quickly and essentially reaches its pre-pandemic level by mid-2020. The over-65 group, by contrast, recovers more slowly and never fully returns to its pre-pandemic baseline — settling around 2.5 speakers per meeting, compared to roughly 3.5 before March 2020. The net result is a persistent narrowing of the age gap in participation.

This convergence is consistent with two (non-mutually-exclusive) interpretations. One is that the initial shift to remote access disproportionately benefited younger, working-age residents — reducing their opportunity costs enough to close the gap with retirees who had always found it relatively easy to attend. The other is that the pandemic itself discouraged older residents from attending, whether due to health risk aversion that persisted beyond the acute phase or a loss of the in-person social routines that had sustained their participation. The DiD design presented in \autoref{sec:did_id} below — which isolates the effect of removing remote access, conditional on the pandemic having already occurred — helps separate these channels.

\subsection{Identification Strategy}\label{sec:did_id}
Due to the lack of variation and clear confounding in the initial introduction of live remote public participation options at the start of the pandemic, we instead exploit the staggered withdrawal of remote public comment across California cities to estimate the causal effect of meeting format on participation. Let $Y_{it}$ be an outcome---total public speakers, mean age, share female, share Democrat---for city $i$ in month $t$. Define $D_{it} = \mathbf{1}\{t \geq G_i\}$, where $G_i$ is the month city $i$ first eliminates live remote public comment. Treatment is absorbing: no city in our sample reintroduces remote access after discontinuing it. We estimate:

\[
Y_{it} \;=\; \alpha_i + \lambda_t + \beta\, D_{it} + \varepsilon_{it},
\]

with city fixed effects $(\alpha_i)$ and month fixed effects $(\lambda_t)$. Under the parallel trends assumption---that, absent a shutoff, outcomes in treated and control cities would have followed similar trajectories---$\beta$ identifies the average treatment effect on the treated (ATT). With staggered treatment timing, standard two-way fixed effects can produce biased estimates due to negative weighting of already-treated units \citep{goodman-bacon_difference--differences_2021, borusyak_revisiting_2024}. We therefore implement the \citet{callaway_difference--differences_2021} estimator, which avoids this issue by restricting comparisons to not-yet-treated and never-treated controls and aggregating group-time ATTs with non-negative weights. Standard errors are clustered at the city level throughout. In order to remove potential confounding due to systematic differences between cities that never introduce remote participation options and those that do, we restrict ourselves to a control set of not-yet-treated units.

The key threat to identification is that cities which eliminated remote access earlier may have been on different participation trajectories than those which kept it. We do not think this is likely. In our survey of city clerks, respondents frequently cited idiosyncratic, administrative reasons for ending remote comment---the expiration of a software contract, changes in how the city attorney interpreted emergency authorizations, or, in several cases, a particularly disruptive ``Zoom bombing'' incident that prompted an immediate return to in-person formats. These accounts suggest that shutoff timing was driven more by logistical happenstance than by strategic considerations about who was participating or what was being discussed.
 
We corroborate this qualitative evidence with three statistical checks. First, we test whether shutoff timing is predicted by observable city characteristics. A regression of remote access duration on pre-pandemic ACS covariates---population, median income, racial composition, educational attainment, renter share, and income inequality---yields an $R^2$ of 0.03, indicating that the decision to withdraw remote comment was largely orthogonal to city demographics. A discrete-time hazard model predicting month-to-month shutoff probability from lagged county COVID case counts and vaccination rates yields a pseudo-$R^2$ of 0.12, with neither predictor statistically significant after controlling for city-level demographic characteristics. The exact timing of shutoffs therefore does not appear to have been driven by local pandemic conditions.
 
Second, we inspect pre-treatment dynamics directly. The event study in \autoref{fig:cs_did_eventstudy_participants} plots dynamic ATTs by event time relative to the shutoff month. Pre-treatment coefficients are flat and jointly indistinguishable from zero, consistent with the parallel trends assumption.
 
Third, we report results both with and without residualizing on meeting-level topic composition shares. If shutoff timing happened to coincide with shifts in agenda content---which could independently affect who shows up---controlling for topic shares should change the estimates. As \autoref{tab:att_zoom_off} shows, it does not: the adjusted and unadjusted estimates are nearly identical.

\subsection{Results}

\begin{figure}[H]
    \centering
    \includegraphics[width=0.75\linewidth]{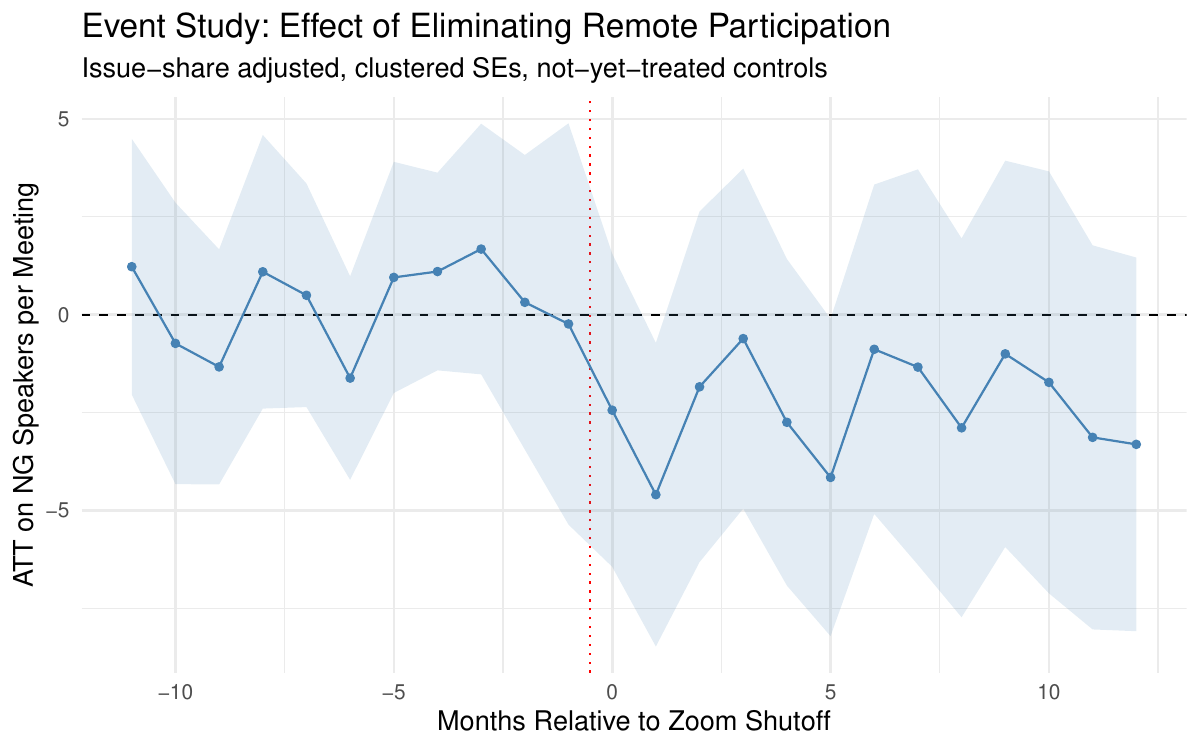}
    \caption{Dynamic treatment effects of eliminating remote public participation}
    \label{fig:cs_did_eventstudy_participants}
\end{figure}

We begin with participation counts, the outcome most directly tied to access costs. The event study in
\autoref{fig:cs_did_eventstudy_participants} plots dynamic ATTs relative to the shutoff month (event time
$e=0$). Pre-treatment coefficients are flat and jointly indistinguishable from zero, supporting the parallel-trends assumption. We find an overall ATT of a reduction of 1.8 speakers ($p=0.33$) following the removal of remote participation options, an effect which increases in magnitude to a reduction of 2.4 speakers ($p=0.21$) when controlling for meeting topical content.

We extend this analysis to the various demographic variables we obtain from matching to the voter record. \autoref{tab:att_zoom_off} summarizes the overall ATT for each outcome. We detect no significant changes the demographic makeup of the participant pool.

\begin{table}[H]
\centering
\caption{Average treatment effects of eliminating remote participation}
\label{tab:att_zoom_off}
\begin{tabular}{lcccccc}
\toprule
 & Pub. Speakers & \% Female & \% Democrat & \% Repeat & Age & \% White \\
\midrule
Baseline & -1.805 & 0.021 & -0.026 & -0.041 & 1.234 & 0.005 \\
 & (1.861) & (0.040) & (0.040) & (0.034) & (1.794) & (0.006) \\

Topic Issue Share & -2.363 & 0.020 & -0.025 & -0.035 & 1.260 & 0.006 \\
 & (1.876) & (0.040) & (0.040) & (0.032) & (1.838) & (0.008) \\

\midrule
$N$ cities & 94 & 94 & 94 & 94 & 94 & 92 \\
$N$ obs & 3,453 & 3,333 & 3,333 & 3,333 & 3,333 & 3,390 \\
Pre-period mean & 16.152 & 0.442 & 0.469 & 0.403 & 54.643 & 0.352 \\
\bottomrule
\end{tabular}
\begin{minipage}{.85\textwidth}
\vspace{4pt}\footnotesize
\textit{Notes:} \citet{callaway_difference--differences_2021}, not-yet-treated controls. 
Standard errors clustered at the town level in parentheses. 
Pre-period mean calculated using 2018--2019 data. 
Topic Issue Share specification residualizes on town$\times$month agenda issue shares. 
$^{*}p<0.10$, $^{**}p<0.05$, $^{***}p<0.01$.
\end{minipage}
\end{table}

\begin{figure}[H]
    \centering
    \includegraphics[width=\linewidth]{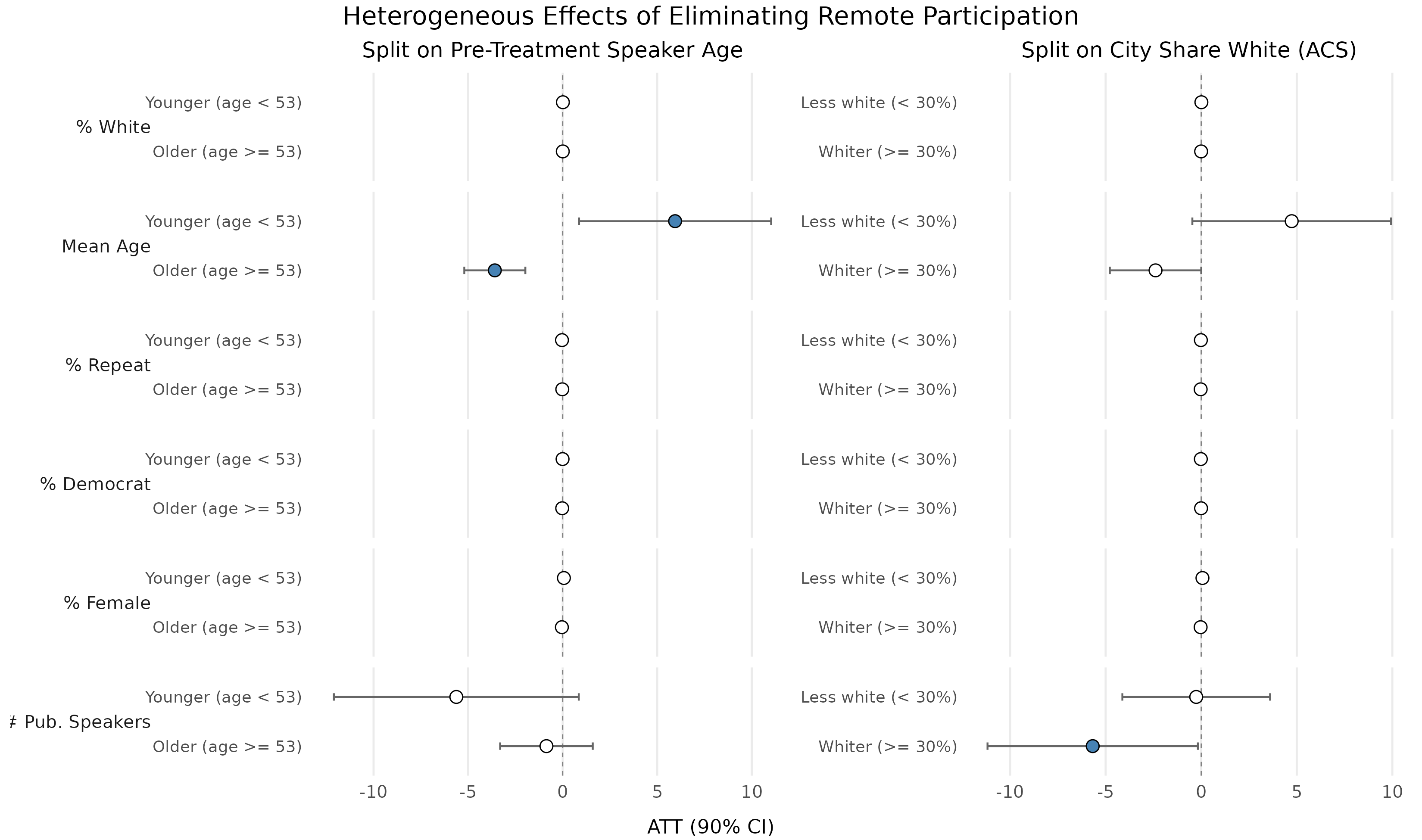}
    \caption{Heterogeneous treatment effects by pre-treatment speaker age and city racial composition}
    \label{fig:het_age_race}
    \begin{minipage}{.85\textwidth}
    \vspace{4pt}\footnotesize
    \textit{Notes:} \citet{callaway_difference--differences_2021} estimator with not-yet-treated controls,
    outcomes residualized on meeting issue topic shares. Points show ATTs; horizontal bars show 95\%
    confidence intervals. Filled points indicate $p < 0.10$. Left panel splits treated cities at the
    median pre-treatment mean speaker age (53 years); right panel splits at the median ACS share white
    (0.30). All never-treated cities serve as controls in both subsamples.
    \end{minipage}
\end{figure}

One possible interpretation is that meeting format simply does not matter much for participation, but the framework in \autoref{sec:framework} also motivates a different reading. If remote access lowers attendance costs for different demographic groups in different cities---older residents facing mobility barriers in some places, younger workers facing scheduling constraints in others---then compositional effects could run in opposite directions across cities and wash out in the pooled sample. We investigate this possibility directly, focusing on age and race as the dimensions along which heterogeneity may be present, motivated by our descriptive analysis in \autoref{sec:who}. We split treated cities at the median of (i) pre-treatment mean speaker age and (ii) city-level share white (from the 2019 ACS), and re-estimate the ATT separately for each subsample. All never-treated cities serve as controls in both splits. \autoref{fig:het_age_race} reports the results.

The age split is striking. In cities with older speaker pools (mean speaker age $\geq 53$), the speaker count does not change significantly ($-0.9$, $p = 0.57$), but mean age \textit{falls} by 3.6 years ($p < 0.001$). Remote access appears to have drawn in older residents---plausibly those facing mobility or health barriers to in-person attendance---who exit when that option is removed, leaving a younger residual pool. In cities with younger speaker pools (mean speaker age $< 53$), the pattern reverses: mean age \textit{rises} by 5.9 years ($p = 0.06$). This is consistent with younger, working-age residents---who may have valued the scheduling flexibility of remote comment---dropping out when in-person attendance is required. These opposing compositional effects cancel in the pooled sample, explaining the null on age in \autoref{tab:att_zoom_off}. The share-white split tells a complementary story about the extensive margin. Eliminating remote access reduces the speaker count by about 5.7 speakers ($p = 0.09$) in whiter cities but has no detectable effect in less-white cities ($-0.3$, $p = 0.91$). The participation decline concentrates entirely in cities whose existing commenter pool most over-represents white residents.

Taken together, these results suggest that the pooled null on participation and demographics conceals meaningful heterogeneity along the very dimensions where public comment is least representative. Remote access expands participation along different margins depending on the city's demographic baseline---but the quantity effect concentrates in whiter cities, precisely those where the existing representation gap is widest. At the same time, even where remote access increases the volume of participation, it does not obviously correct the underlying compositional skew documented in \autoref{sec:who}. Lowering the cost of attendance brings in more speakers, but not necessarily more representative ones.

\section{Conclusion}
Local government meetings remain the most durable and visible arena of participatory democracy in the United States. Yet, they have remained chronically under-studied, with most existing evidence suffering from small sample sizes and inconsistent coverage. In this paper, we compile what is to our knowledge the largest, most comprehensive dataset of city council meeting transcripts in a given region of the United States, transcribing and analyzing over 25,000 meetings across 115 cities in California over the last decade. By linking this data to voter and property ownership records, we are able to provide a highly granular and detailed look at the structure and format of these meetings, characterize who participates and how participation changes with meeting content, and assess the causal effect of changing barriers to meeting access (in the form of remote participation options) on participation rates and demographics. 

Our evidence shows that meetings are long, frequent, have many participants, and cover a variety of topics; participants tend to be older, whiter, more male, more liberal, and more likely to be homeowners than the full registered voter population; participation patterns vary across cities depending on city size, median income, level of education, renter share, development, and racial diversity; participants are drawn to meetings that focus more on Land Use \& Zoning, Housing \& Homelessness, and Social Equity; and increasing costs of attendance by removing a remote participation option leads to a short-term reduction in the number of public participants but does not clearly change the demographic composition of speakers.

This project also opens up compelling avenues for future work. This includes examining how measures discussed in meetings evolve through the meeting process; how public comment sways future vote outcomes; and how comments and votes by councilmembers are correlated with electoral outcomes. More broadly, this line of research shows significant promise for deepening our understanding of local government dynamics and informing the design of more effective, representative democratic institutions.

\newpage

\bibliography{references}

\newpage

\appendix

\section*{Appendix}

\section{LLM Prompts}\label{apx:llm_prompts}

\subsection{Speaker Identification}
\small
\begin{verbatim}
You are analyzing a transcript of a city council meeting.

Below is a chronological transcript produced by an automatic speech \
diarization system. The diarization model segments the audio into chunks \
and estimates which speaker(s) are talking in each chunk.

**How to read the transcript format:**

- When the model is confident that a single speaker accounts for >=90%
segment, the line looks like:
  SPEAKER_06 (00:01:23 - 00:02:45):  [their speech text]

- When no single speaker dominates a segment (i.e., multiple speakers are \
detected), the line shows each speaker's estimated share:
  {{SPEAKER_06: 0.70, SPEAKER_00: 0.30}} (00:00:00 - 00:00:07):  [text]
  This means SPEAKER_06 spoke roughly 70%
spoke roughly 30%
to specific speakers is uncertain.

**Important notes on diarization quality:**

- The diarization is imperfect. It is possible (and common) that two \
different speaker IDs actually refer to the same real person. Use context \
clues (e.g., the speaker introduces themselves, or another speaker addresses \
them by name) to determine when two IDs are the same person. When you \
believe two or more speaker IDs are the same person, assign them the same \
name.
- Conversely, the same speaker ID usually does refer to one person, but \
diarization errors can occasionally assign one ID to multiple people.

The transcript includes both government officials (council members, mayor, \
city clerk, city manager, city attorney, city planner, police, fire, and \
any other city staff) and non-government speakers (members of the public, \
community organizations, businesses, etc.). Use the full transcript for \
context — for example, a city clerk may announce the next speaker's name \
before they speak.

Your task: for each of the {n_speakers} speaker IDs listed below, return a \
JSON object with the following three fields:

1. "name": Your best guess for the speaker's real name, based ONLY on what \
   appears in the transcript (e.g., a speaker introducing themselves, or \
   another speaker addressing them by name).
   - If you cannot confidently identify a name, return "NA".
   - If you can only identify a first name or last name, return only that — \
     do NOT fabricate or guess the missing part. For example, if a speaker \
     says "Hi, I'm Maria" and no last name is ever mentioned, return \
     "Maria", not "Maria [Guess]".
   - Do NOT invent names. Only return names that are explicitly stated or \
     very clearly implied in the transcript text. 
   - If the transcript only contains an individual's first name and last initial, please \ 
     just return the first name and last name initial without guessing at \
     the full last name (e.g., "Olivia M." should remain "Olivia M.")
   - If you believe two speaker IDs are the same person, give them the same \
     name.
   - Do not include titles like "Councilmember", "Pastor", etc.. For example, "Councilmember \
   Olivia Martin" should be returned as "Olivia Martin"

2. "gov": Whether the speaker is part of the city government (labeled "G") \
   or a member of the public (labeled "NG"). City government includes \
   council members, mayor, city clerk, city manager, city attorney, city \
   planner, police, fire, and any other city staff.

3. "group": The capacity in which the speaker is appearing before the \
   council. Use exactly one of the following single-letter codes:
   - "I": individual member of the public (not representing any group)
   - "L": labor union
   - "A": advocacy group
   - "E": environmental group
   - "B": business
   - "O": other organized group not covered above
   - "NA": speaker is a government official (gov = "G")

Return ONLY a valid JSON object mapping each speaker ID to an object with \
these three fields. No commentary, explanation, or code — just JSON.

Speaker IDs to identify ({n_speakers} total):
{speaker_list}

Here is the full transcript:

{transcript}
\end{verbatim}

\subsection{Voting and Issue Identification}

\begin{verbatim}
    You are analyzing a transcript of a California city council meeting. The \
transcript was produced by an automatic speech diarization system and is \
formatted chronologically. Each speaker block is labeled with a speaker ID \
and the timestamps of that segment in the audio recording:
  SPEAKER_06 (00:01:23 - 00:02:45):  [speech text]
Some segments show multiple speakers with estimated fractions:
  {{SPEAKER_06: 0.70, SPEAKER_00: 0.30}} (00:00:05 - 00:00:12):  [speech text]

Analyze this transcript and identify the key \
issues/topics that the council weighed in on (i.e., received council member \
comment or came up for a vote). Omit issues that do not meet this threshold.

For each issue, provide 9 fields:

1) "issue": A short descriptive title. Always include a specific street \
address, parcel number, or project name if one is mentioned in the \
transcript (e.g., "1075 Pomeroy Avenue Rezoning", "Santana West Mixed-Use \
Project", "Whitmore Ranch Annexation"). Favor substantive names over \
procedural ones — use "Parking Rate Changes" not "Agenda Item B".

2) "summary": A concise description (~70 characters) of what is under \
consideration. Include any specific address or project name. Do NOT begin \
with filler like "The city council considered" — lead with substance. \
Include the procedural posture if discernible from the transcript (e.g., \
"Second reading of ordinance to rezone...", "Appeal of PC denial of..."). \
Do NOT include the outcome of any vote. \
Good examples: \
"Rezone 1075 Pomeroy Ave from R3-18D to PD for 5 homes", \
"Second reading: rezone Berryessa flea market site (61.5 acres)", \
"Appeal of PC denial of rezone at 350 Central Ave to allow 70 homes", \
"Ordinance introduction: rezone 2481 Deerwood Dr to high-density residential". \
Bad examples: \
"The city council considered a rezoning request", \
"A rezone was discussed and voted on".

3) "public": true/false — was there any public comment or testimony?

4) "vote": true/false — was a vote taken?

5) "vote_res": The vote tally as "yea-nay-abstain" (e.g., "3-2-0"). \
Always report votes in favor first, then opposed, then abstentions. \
Do not report absences. If no vote was held, report "None".

6) "vote_outcome": A short phrase describing what the vote decided — \
the practical result for the item. Focus on what happened to the item, not \
the tally. Examples: "Approved", "Denied", "Approved on consent calendar", \
"Continued to [date]", "Continued indefinitely", "Introduced on first \
reading", "Adopted on second reading", "Referred to Planning Commission", \
"Tabled", "Withdrawn", "Approved with conditions", "Rescission approved", \
"Appeal denied (PC decision upheld)", "Direction given to staff", \
"Motion failed for lack of second". If the transcript is ambiguous about \
what the vote accomplished, write "Unclear". If no vote, write "None".

7) "vote_stage": Classify the vote's procedural significance:
   - "final": a dispositive action that resolved the matter (approval, \
     denial, adoption on final reading, withdrawal).
   - "procedural": advanced or delayed the item without final resolution \
     (first reading/introduction, continuance, referral to committee, \
     tabling, motion to reconsider, direction to staff).
   - "unclear": the transcript does not make the stage clear.
   - "none": no vote was held.

8) "timestamp_start": The start time of the earliest speaker block that \
covers discussion of this issue. The transcript labels each speaker block \
with a time range in the format "SPEAKER_XX (HH:MM:SS - HH:MM:SS):"; \
use the HH:MM:SS start time from the first relevant block. Include any \
speech related to debate, public comment, or the voting portion — the \
range should cover the entirety of relevant conversation. Report as a \
string in "HH:MM:SS" format. Use null if the transcript does not include \
timestamps.

9) "timestamp_end": The end time of the latest speaker block that covers \
discussion of this issue. Use the HH:MM:SS end time from the \
"SPEAKER_XX (HH:MM:SS - HH:MM:SS):" label of the last relevant block, \
spanning the full conversation including debate, public comment, and \
voting. Report as a string in "HH:MM:SS" format. Use null if the \
transcript does not include timestamps.

Format your response as a JSON array of objects with exactly these 9 keys: \
"issue", "summary", "public", "vote", "vote_res", "vote_outcome", \
"vote_stage", "timestamp_start", "timestamp_end".

Here is the transcript you should analyze:

{transcript}
\end{verbatim}

\subsection{Issue Topic Discovery}

\begin{verbatim}
You are an expert text-classification model with a focus on city governance. Below is a dictionary of 
summaries of issues from city council meetings: each key is an ID for each summary, which is 
stored as the corresponding value. Note that the ID number assigned to each issue is arbitrary 
and contains no information. Your task is to learn no fewer than 5 and no more than 10 topics 
that collectively span the set of issues contained in this list.

Return a name for each topic, a description of what it encompasses, and 5 representative issue 
summaries from the set I will provide to you that are very closely associated with the given topic.

Return only your discovered topics and associated descriptions and examples. Do not return any code.

Here is the dictionary of issue summaries, mapping issue ID to the one-sentence summary:

{issue_dict}
\end{verbatim}

\subsection{Issue Topic Unification}

\begin{verbatim}
You are an expert text-classification model specializing in city governance. Your task is to 
unify the topics identified from different chunks of city council issues into exactly 10 topics.

Below is a dictionary of topics from {len(topic_chunks)} chunks of a larger dataset. Each key in 
this dictionary corresponds to one chunk, and each value is a list of topics, each with its own 
title, description, and representative examples:

{topic_chunks}

**Instructions:**

1. Combine and unify these topics into exactly 10 final topics. 
If multiple topics overlap or are extremely similar, merge them.
Choose concise and representative titles for the final set.

2. Provide a brief, 1-3 sentence description for each of the 10 final topics.

3. Include exactly 5 representative examples per topic. 
Select the examples from among all the examples in the provided topic chunks.

4. Output the final topics in the following JSON structure:

[
  {{
    "topicID": 0,
    "topicTitle": "Your Topic Title",
    "description": "Short description (1-3 sentences)",
    "representativeExamples": [
      "Example 1",
      "Example 2",
      "Example 3",
      "Example 4",
      "Example 5"
    ]
  }}
]
Do not return any code. Return ONLY valid JSON.
\end{verbatim}

\subsection{Topic Classification}

\begin{verbatim}
You are an expert text-classification model specializing in city governance.
Below is a list of discovered topics that commonly occur in city council meetings.
Each topic has an ID number (0-9), a title, a description, and 5 representative examples:

{topics_json}

Below is a dictionary mapping issue IDs (integers) to issue summary texts from city council meetings.
Classify each issue into exactly one of the 10 topics above.

Return ONLY a valid JSON object mapping each issue ID (as a string) to its topic ID (as an integer).
Do not include any explanation, markdown, or extra text.

Here is the dictionary of issue summaries:

{issue_summaries}
\end{verbatim}

\subsection{Land Use Pro- vs. Anti-Development Scores}

\begin{verbatim}
You are given:
- Issue title: {issue}
- Issue summary: {summary}

Below is a dictionary where each key is a unique identifier for a comment on THIS issue,
and each value is the transcribed speech text for that comment.

Your task is to analyze each value (comment text) and predict whether the speaker is in favor of
or opposed to a more development/growth outcome on this issue on a continuous scale of 1 to -1 where:
-  1.0 = clearly supportive of more development/growth (e.g., favors approval of a project,
          wants more development, etc.)
-  0.0 = neutral / unclear
- -1.0 = clearly opposed to more development/growth (e.g., wants more restrictions on new
          development, has a lot of concerns about development)

Return **only** a list of lists, where each inner list has exactly two elements:
[unique_id, score]
- unique_id must be a string
- score must be a float in [-1, 1]

Do **not** include any explanation or extra text. The very first characters must be [[

Here is the input dictionary:
{input_dictionary}
\end{verbatim}
\normalsize

\section{L2 Matching Algorithm}\label{apx:l2_match_algo}
This section provides an overview of the algorithm used to match names identified by the LLM from meeting transcripts to the L2 voter record dataset. The algorithm operates as follows:

\begin{enumerate}
    \item Select a meeting and its corresponding extracted LLM-identified speaker names, subsetting to speakers tagged by the LLM as members of the public
    \item Pull down the subset of L2 data corresponding to that town and year the selected meeting occurred
    \item Select a name from the LLM-identified set and ensure that the name appears in the underlying meeting transcript (to filter out hallucinations)
    \begin{itemize}
        \item If both a first and last name are provided, ensure that both the first name and last name appear in the transcript, but not necessarily together
        \item If both a first and last name are provided, but only the first name appears in the transcript, truncate the last name to the initial only
        \item If a first name and last initial are provided, ensure that both appear together (consecutively) in the underlying transcript
    \end{itemize}
    \item Compare the name to all registered voters in L2 and assign a match priority based on closeness of match, based on the following logic:
    \begin{itemize}
        \item If both a first and last name are provided, use the following match priority order:
        \begin{itemize}
            \item Exact match on first name and last name
            \item Exact match on last name and either exact first name nickname match or metaphone match for first name
            \item Exact match on first name and metaphone match on last name
            \item Exact first name nickname match and metaphone match on last name
            \item Exact match from LLM-identified first name to L2 middle name and exact last name match
            \item Exact match from LLM-identified first name nickname to L2 middle name and exact last name match
            \item Exact match from LLM-identified first name to L2 middle name and metaphone match on last name
            \item Exact match from LLM-identified first name nickname to L2 middle name and metaphone match on last name
            \item Metaphone match on first name and metaphone match on last name
            \item Metaphone match on LLM-identified first name to L2 middle name and exact match on last name
            \item Metaphone match on LLM-identified first name to L2 middle name and metaphone match on last name
            \item Exact match or metaphone match on last name only
        \end{itemize}
        \item If only a first name and last initial are provided, use the following priority match order (assuming in all cases that the last initial matches exactly):
        \begin{itemize}
            \item Exact first name match
            \item Exact first name nickname match
            \item Exact match from LLM-identified first name to L2 middle name
            \item Exact match from LLM-identified first name nickname to L2 middle name
            \item Metaphone match on first name
            \item Metaphone match on LLM-identified first name to L2 middle name
        \end{itemize}
    \end{itemize}
    \item Consider only candidate matches in the class with highest match priority and apply the following logic: 
    \begin{itemize}
        \item If all priority sets are empty, do not match the name at all.
        \item If there is a single matched name in the highest priority class, assign it as the matched name.
        \item If there is more than one matched name in the highest priority class and the LLM did not provide a complete last name or we had no match (either exact or metaphone) on the given name (including potential nicknames/middle names), do not match the name at all.
        \item If there is more than one matched name in the highest priority class and the LLM provided a complete last name and we had a match (either exact or metaphone) on the given name (including potential nicknames/middle names), assign the matched name in L2 with the greatest weighted Jaro-Winkler similarity (assigning 80\% weight to last name and 20\% to first name), after filtering out candidate matches which do not have an exact match on the first initial of the last name.
    \end{itemize}
    \item Repeat Steps 3-5 for all names in the given meeting.
    \item Repeat Steps 1-6 for all meetings in the corpus.
\end{enumerate}

If this process fails to yield a match for a given speakers, we attempt to instead find a match in the preceding or subsequent year of L2 data. This allows us to account for people who move to or from a given town, or change their registered voter address, in the middle of the year, before or after the L2 record collection cutoff for that year. If we find a unique match for the given (previously unmatched) speaker in either the preceding or subsequent year of L2 data for the given town, we assign it as the matched name. If both the preceding and subsequent years of L2 yield conflicting matches, we assign the match with the higher match priority, as outlined in Step 4 above. If both the preceding and subsequent years of L2 yield conflicting matches and have identical match priority, we leave the speaker unmatched.

\section{LLM Validation}\label{apx:llm_name_valid}

\subsection{Name and Status Extraction}

To assess the accuracy with which the LLM correctly identifies speakers as members of the public and attributes speaker text to a name, we employed two research assistants to create a hand-labeled set of speaker IDs matched to names and governmental status. The RAs were provided with 62 randomly sampled, stratified meeting transcripts formatted as ``SPEAKER 01: TEXT; SPEAKER 02: TEXT.'' Their task was to guess based on the transcript context the name and government status for each speaker.\footnote{In the initial version of this exercise performed in September 2025, our first RA only completed this task for the top 25 speakers in each meeting by speaking time. We therefore cannot use those transcripts to compute meeting-level precision and recall and instead use a new human-validated sample for those calculations.} The result is a dataset of 1,300 labeled speakers.

\subsubsection{Government Status}

As shown in \autoref{tab:gov_confusion}, our RAs agree with the LLM assignment of government status 87\% of the time. The LLM was more likely to disagree with the RA in labeling speakers as G when the RA labeled the speaker as NG. Reviewing ten of the 91 cases of disagreement of this type between the LLM and the RA, we could find no example of a clear mistake on the part of the LLM: the speakers plausibly seemed to be city staff (though not on the city council) making presentations. 

\begin{table}[htbp]
\caption{Government status confusion matrix: RA ground-truth labels vs.\ LLM predictions. }
\label{tab:gov_confusion}
\centering
\begin{tabular}{lcc|c}
\toprule
 & {LLM: G} & {LLM: NG} & {Total} \\
\midrule
{RA: G} & 676 & 46 & 722 \\
{RA: NG} & 91 & 454 & 545 \\
{RA: NA} & 18 & 15 & 33 \\
\midrule
{Total } & 785 & 515 & 1,300 \\
\bottomrule
\end{tabular}
\end{table}

\subsubsection{Name}

Name extraction is inherently harder than government status classification because transcripts often contain only partial names, phonetic misspellings, or no spoken name at all. \autoref{tab:none_confusion} cross-tabulates whether the RA and LLM each identified a name for the 1,300 labeled speakers.

\begin{table}[H]
\caption{Confusion matrix: RA vs.\ LLM identification of a name. Ten RA labels containing multiple names (slash/semicolon separated) were treated as NONE.}
\label{tab:none_confusion}
\centering
\begin{tabular}{lccc}
\toprule
 & LLM $\neq$ NONE & LLM = NONE & Total \\
\midrule
RA $\neq$ NONE & 960 & 74 & 1,034 \\
RA = NONE & 128 & 138 & 266 \\
\midrule
Total & 1,088 & 212 & 1,300 \\
\bottomrule
\end{tabular}
\end{table}

Both sources identified some form of name for 960 speakers; both returned no name for 138. The asymmetric disagreements---128 where only the LLM produced a name and 74 where only the RA did---are the cases of primary concern.

The 128 LLM-only names are potential hallucinations. Of these, 36 are single-word (e.g., a first name only), which our downstream matching algorithm (detailed in \autoref{apx:l2_match_algo}) would not link to the L2 voter file regardless. Of the 92 multi-word names, only 22 (24\%) matched an L2 voter record, suggesting that most LLM-only names do not propagate into downstream demographic analyses. The 74 RA-only names represent cases where the LLM was more conservative; because these speakers simply receive no demographic match, they reduce recall but do not introduce bias.

Among the 960 speakers where both sources identified a name, \autoref{tab:name_match_both_named} reports agreement rates.

\begin{table}[H]
\caption{Name matching accuracy among speakers where both RA and LLM identified a name. Name similarity uses token-sort ratio (0--100); last-name similarity uses character-level ratio (0--100).}
\label{tab:name_match_both_named}
\centering
\begin{tabular}{lc}
\toprule
Metric & Value \\
\midrule
Exact match & 58.3\% (560/960) \\
Fuzzy match ($\geq$70) & 73.1\% (702/960) \\
Mean name similarity & 82.4 \\
Median name similarity & 100.0 \\
\midrule
Last name exact match & 73.8\% (708/960) \\
Mean last-name similarity & 84.8 \\
Median last-name similarity & 100.0 \\
\bottomrule
\end{tabular}
\end{table}

Exact full-name matches occur 58\% of the time, rising to 73\% at a fuzzy threshold of 70 (token-sort ratio). Last-name exact match is higher at 74\%, reflecting the fact that many disagreements stem from one source providing a full name where the other provides only a surname. The median similarity score of 100 confirms that the majority of pairs agree perfectly, with a long left tail driving the mean down.

\autoref{tab:name_mismatch_diagnosis} diagnoses the 258 pairs that fell below the fuzzy threshold.

\begin{table}[H]
\caption{Diagnosis of both-named speakers that failed fuzzy matching (token-sort ratio $<$ 70). Categories applied in order (first match wins).}
\label{tab:name_mismatch_diagnosis}
\centering
\begin{tabular}{lrrp{5.5cm}}
\toprule
Category & Count & \% & Example Pairs (RA / LLM) \\
\midrule
RA one name only & 85 & 32.9 & Rocha / Orlando Rocha; Sandoval / Anthony Sandoval \\
LLM one name only & 16 & 6.2 & Laura McDowell / Laura; Steve Carell / Steve \\
Last name match, first name differs & 2 & 0.8 & Maran Frometa / Claudia Frometa \\
Spelling variant & 30 & 11.6 & Sornio / Andrew Osornio; Alvern / Alvord \\
Completely different (possible misattribution) & 125 & 48.4 & Barry Gilbert / Jordan Brandman; Dan Orgel / Paul Emery \\
\midrule
Total & 258 & 100.0\% & \\
\bottomrule
\end{tabular}
\end{table}

The single largest category (48\%, $n=125$) comprises completely different names. However, this largely reflects misattribution rather than hallucination: in 86\% of these cases, the LLM's name appears elsewhere in the transcript, indicating that the LLM assigned a real speaker's name to the wrong segment. An additional 33\% of mismatches arise because the RA recorded only a surname while the LLM recovered a full name---in 56 of these 85 cases, the LLM's additional name component appears verbatim in the transcript.

Finally, \autoref{tab:precision_recall} reports set-level precision and recall, collapsing within each meeting to unique name sets.

\begin{table}[H]
\caption{Set-level name precision and recall}
\label{tab:precision_recall}
\centering
\begin{tabular}{lcc}
\toprule
Metric & Full name & Last name \\
\midrule
Precision & 68.5\% (146/213) & 77.5\% (165/213) \\
Recall & 75.5\% (145/192) & 84.4\% (162/192) \\
\bottomrule
\end{tabular}
\begin{minipage}{.85\textwidth}
\vspace{4pt}\footnotesize
\textit{Notes:} For each meeting, we collect the set of unique names identified by the RA and by the LLM. A name counts as matched if any name in the other set exceeds the fuzzy threshold ($\geq$ 70). Full-name matching uses token-sort ratio; last-name matching uses character-level ratio on the final word only.
\end{minipage}
\end{table}

Last-name precision is 78\% and recall is 84\%, confirming that the LLM captures most speakers who appear in a meeting while introducing a manageable rate of names beyond those which the RA could identify.

Taken together, these results suggest that name extraction errors are predominantly misattributions across speaker segments rather than fabricated names, and that the most consequential errors---false names that match to voter records and could distort demographic analyses---are rare.

\subsection{Vote and Issue Identification}

Validating issue and vote identification is a substantially more demanding task than validating speaker names or government status. Rather than classifying pre-segmented speaker turns, the labeler must read each transcript line by line---identifying where one issue ends and another begins, distinguishing agendized items from unagendized public comment, and tracking whether and how the council voted on each item. The seven meetings in our validation sample total approximately 360 pages of transcript text, ranging from 2 pages (Santa Clara) to over 90 pages (Santa Barbara), requiring close reading of the full document to produce ground-truth labels.

To construct these labels, we employed a research assistant to independently label the seven meetings, drawn from the same stratified sample used in the speaker identification exercise. The RA received the identical prompt given to the LLM (detailed in \autoref{apx:llm_prompts}) and manually read each transcript, recording the following for each identified issue: title, summary, whether it was agendized, whether a vote occurred, vote tally, and timestamps.\footnote{We analyze one meeting each from seven cities: Riverside (2018-02-27), Compton (2018-02-13), Victorville (2015-06-16), Santa Barbara (2020-03-10), Paramount (2020-03-17), Santa Clara (2019-12-09), and Turlock (2018-03-13). The Santa Barbara meeting was split across four separate video/transcript files on that date.}

\subsubsection{Issue Detection}

We match RA-labeled issues to LLM-labeled issues using a combined score of timestamp intersection-over-union (IoU, weighted 0.7) and fuzzy title similarity (weighted 0.3), with a greedy best-first algorithm requiring either IoU $\geq$ 0.3 or title similarity $\geq$ 60.\footnote{For RA interval $[s_r,e_r]$ and LLM interval $[s_\ell,e_\ell]$, we compute timestamp overlap as $\mathrm{IoU}=\frac{\max\{0,\min(e_r,e_\ell)-\max(s_r,s_\ell)\}}{\max(e_r,e_\ell)-\min(s_r,s_\ell)}$. Title similarity is the RapidFuzz token-sort ratio, rescaled to $[0,1]$. The combined match score is $0.7\times \mathrm{IoU} + 0.3\times \mathrm{TitleSim}$. Within each meeting, we score all RA--LLM pairs, sort them from highest to lowest score, and greedily accept one-to-one matches, provided the pair satisfies $\mathrm{IoU}\geq 0.3$ or title similarity $\geq 60$.} \autoref{tab:vote_issue_summary} reports the results. The RA and LLM identified similar numbers of issues overall. Among matched pairs, timestamp overlap is high, indicating that matched issues refer to the same segment of the meeting.

\begin{table}[htbp]
\caption{Vote/issue identification validation summary}
\centering
\begin{tabular}{lr}
\toprule
Metric & Value \\
\midrule
Meetings validated & 7 \\
RA issues identified & 133 \\
LLM issues identified & 131 \\
Matched pairs & 102 \\
RA-only (LLM missed) & 31 \\
LLM-only (LLM extra) & 29 \\
\midrule
Issue recall (matched / RA total) & 76.7\% \\
Issue precision (matched / LLM total) & 77.9\% \\
\midrule
Mean timestamp IoU (matched pairs) & 0.93 \\
Median timestamp IoU (matched pairs) & 0.98 \\
\bottomrule
\end{tabular}
\label{tab:vote_issue_summary}
\end{table}

The raw precision and recall figures understate agreement for two reasons. First, one of the seven meetings (Santa Barbara, 2020-03-10) was split across four separate video/transcript files. We verified that all 18 LLM-only issues attributed to Santa Barbara appear in those other Santa Barbara transcripts from the same day rather than reflecting hallucinated issues. Excluding the Santa Barbara meeting entirely, issue-level precision rises to 89.1\% (90/101) while recall is essentially unchanged at 76.3\% (90/118).

Second, the 28 RA-only issues (excluding Santa Barbara) fall into two categories that explain the recall gap. Of the 10 that were agendized, several are routine procedural items---including consent calendar approvals, closed-session requests or reports, and a warrant approval---that the LLM tends to consolidate or omit. The remaining 18 are unagendized, and the majority reflect a granularity difference in how the RA and LLM delimit ``issues'' during public comment periods. For example, in Compton---which accounts for half of all RA-only items---the RA recorded each individual public commenter as a separate issue (e.g., ``member of the public requests a job in the city,'' ``member of the public asks for help locating his missing daughter''), while the LLM grouped these into broader topics. Neither treatment is incorrect, but the RA's finer-grained approach inflates the denominator for recall. Because our downstream analyses operate at the level of substantive policy issues rather than individual public comments or procedural motions, this granularity difference is unlikely to affect our findings.

\subsubsection{Agendized Classification}

Among the 102 matched issue pairs, the RA and LLM agree on whether the issue was agendized 97.1\% of the time (99/102). The three disagreements are attributable to ambiguous framing rather than clear errors: a land use appeal in Victorville that the RA coded as unagendized but the LLM coded as agendized (it was a formal public hearing), a COVID-19 emergency ratification in Paramount, and a library update in Turlock.

\subsubsection{Vote Detection and Tally}

Among the 101 matched pairs with non-missing vote indicators, the RA and LLM agree on whether a vote occurred 97.0\% of the time (98/101). \autoref{tab:vote_confusion_matched} shows the confusion matrix.

\begin{table}[htbp]
\caption{Vote detection confusion matrix among matched issue pairs (1 pair excluded due to missing values)}
\centering
\begin{tabular}{lcc|c}
\toprule
 & {LLM: Vote} & {LLM: No Vote} & {Total} \\
\midrule
{RA: Vote} & 50 & 1 & 51 \\
{RA: No Vote} & 2 & 48 & 50 \\
\midrule
{Total} & 52 & 49 & 101 \\
\bottomrule
\end{tabular}
\label{tab:vote_confusion_matched}
\end{table}

Among the 50 matched pairs where both the RA and LLM recorded a vote, the vote tally (\texttt{vote\_res}) agrees exactly 96.0\% of the time (48/50). One disagreement reflects column-order ambiguity rather than a counting error: in Compton, the RA recorded 4-0-0 while the LLM recorded 0-4-0 for the same four-vote action. For the other disagreement, we checked the official Santa Barbara minutes for the Alameda Padre Serra appeal; those minutes record four ayes and three noes, so the LLM's 4-3-0 coding is correct and the RA's 4-0-3 coding is not.\footnote{The official minutes for the March 10, 2020 Santa Barbara City Council meeting record a majority roll call vote on the Alameda Padre Serra appeal with four ayes and three noes. See \citet{city_of_santa_barbara_city_2020}. We were unable to locate the meeting minutes for the relevant Compton City Council meeting.}

Vote stage agreement is 90.0\% (45/50). The five disagreements typically involve the RA coding a vote as ``unclear'' where the LLM coded it as ``final,'' reflecting the difficulty of distinguishing procedural from final votes in transcript text.

\subsubsection{Summary}

\autoref{tab:vote_issue_counts_by_meeting} reports per-meeting issue counts.

\begin{table}[htbp]
\caption{Issue counts by meeting}
\centering
\begin{tabular}{llrrrr}
\toprule
Town & Date & RA & LLM & Matched & RA-only \\
\midrule
Compton & 2018-02-13 & 35 & 26 & 21 & 14 \\
Paramount & 2020-03-17 & 15 & 14 & 11 & 4 \\
Riverside & 2018-02-27 & 13 & 12 & 12 & 1 \\
Santa Barbara & 2020-03-10 & 15 & 30 & 12 & 3 \\
Santa Clara & 2019-12-09 & 2 & 2 & 2 & 0 \\
Turlock & 2018-03-13 & 17 & 15 & 13 & 4 \\
Victorville & 2015-06-16 & 36 & 32 & 31 & 5 \\
\midrule
{Total} & & 133 & 131 & 102 & 31 \\
\bottomrule
\end{tabular}
\label{tab:vote_issue_counts_by_meeting}
\begin{minipage}{.85\textwidth}
\vspace{4pt}\footnotesize
\textit{Notes:} Santa Barbara's elevated LLM count reflects the multi-video split discussed in the text. Compton's elevated RA-only count reflects several procedural items (warrants, closed-session requests) that the LLM consolidated or omitted.
\end{minipage}
\end{table}

Taken together, these results suggest that the LLM reliably identifies the substantive issues discussed in city council meetings and accurately classifies whether they were agendized and whether a vote was taken. The primary source of disagreement is the treatment of routine procedural items---consent calendars, closed-session announcements, and warrant approvals---which the LLM tends to omit or consolidate. Because our downstream analyses focus on substantive policy issues rather than procedural votes, this pattern is unlikely to bias our findings. Conditional on detecting an issue, the LLM's vote tallies are highly accurate, with the two observed disagreements reflecting ambiguity in the yea-nay-abstain column ordering rather than miscounts.

\subsubsection{Summary}

Taken together, these results suggest that the LLM reliably identifies the substantive issues discussed in city council meetings and accurately classifies whether they were agendized and whether a vote was taken. The primary source of disagreement is the treatment of routine procedural items---consent calendars, closed-session announcements, and warrant approvals---which the LLM tends to omit or consolidate. Because our downstream analyses focus on substantive policy issues rather than procedural votes, this pattern is unlikely to bias our findings. Conditional on detecting an issue, the LLM's vote tallies are highly accurate, with the two observed disagreements reflecting ambiguity in the yea-nay-abstain column ordering rather than miscounts.

\section{Additional Tables \& Figures}\label{apx:addl}

\begin{figure}[H]
    \centering
    \includegraphics[width=0.6\linewidth]{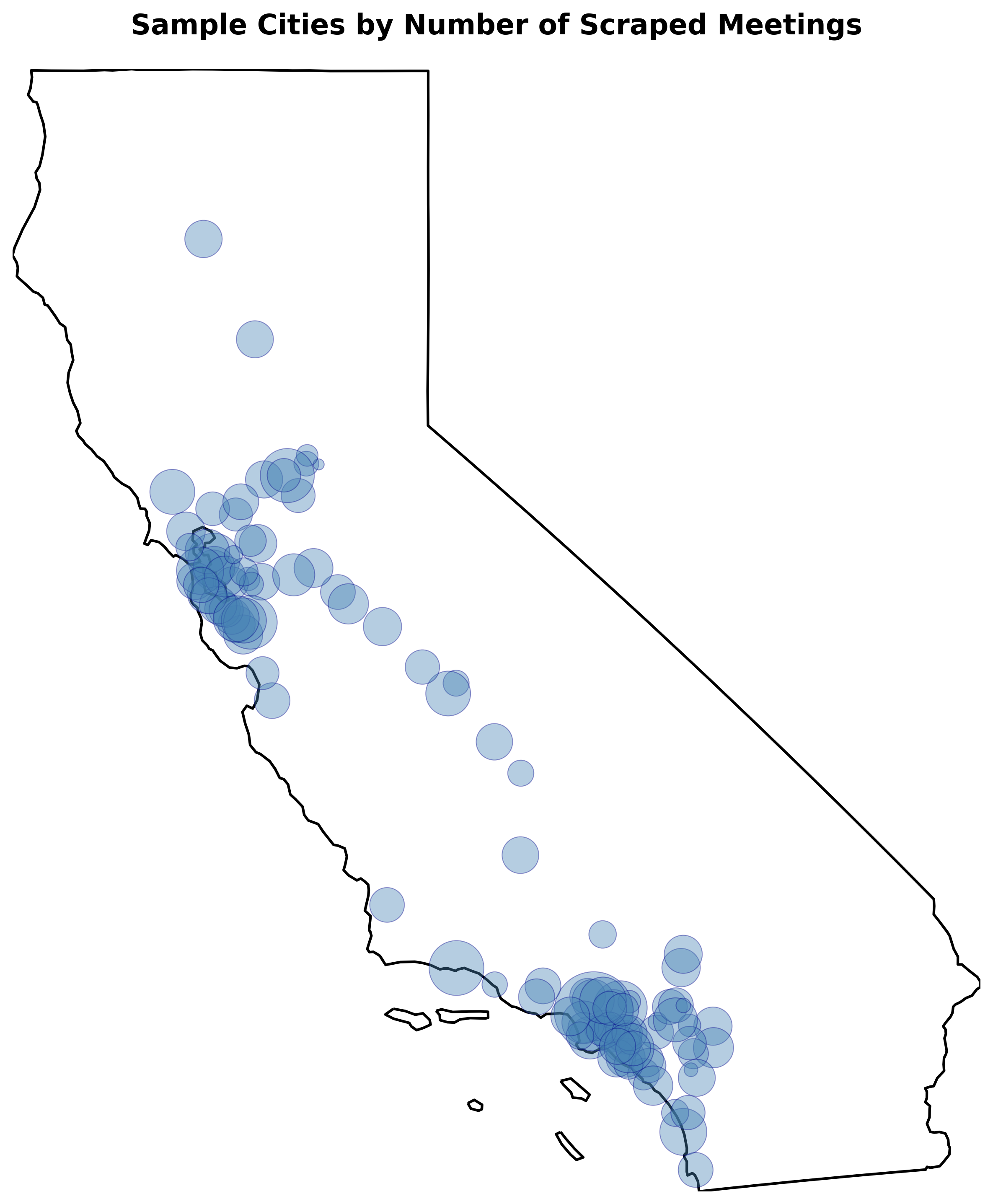}
    \caption{Geographic coverage of collected data. Dots reflect sampled cities, with size of the dot corresponding to the number of meetings contained in our dataset.}
    \label{fig:map}
\end{figure}

\begin{table}[H]
\caption{Breakdown of towns in our sample by channel through which city council meetings are made available to the public. Population estimates provided by \citet{state_of_california_department_of_finance_e-4_2025}.}
\label{tab:channel_coverage}
\centering
\begin{tabular}{lccc}
\toprule
{Channel} & {\# Towns} & {Total Population} & {\% of California Population}\\
\midrule
YouTube & 52 & 9{,}346{,}755 & 23.6\\
Other & 63 & 11{,}213{,}856 & 28.3\\
\midrule
Total & 115 & 20{,}560{,}611 & $52.0$ \\
\bottomrule
\end{tabular}
\vspace{0.5em}

\end{table}

\begin{table}[H]
\centering
\caption{Examples of Meeting Topics by Category}
\label{tab:topic_issue_examples}
\renewcommand{\arraystretch}{1.15}
\begin{tabular}{p{4.5cm}p{12.5cm}}
\toprule
\textbf{Category} & \textbf{Example Topics} \\
\midrule
Governance \& Admin & Consent Calendar; League of California Cities Voting Delegate; Planning Commission Appointments; Approval of Minutes \\
Land Use \& Zoning & Short-Term Rental Regulations; Accessory Dwelling Unit (ADU) Ordinance; California Building Standards Code Adoption; Noriega Road and Old Farm Road Residential Project \\
Infrast. \& Transport. & National Public Works Week Proclamation; Urban Water Management Plan; State Legislative Package and Caltrans Landscaping; Municipal Fiber Optic Infrastructure \\
Community Services & Parks and Recreation Month Proclamation; National Library Week Proclamation; Older Americans Month Proclamation; Parks Make Life Better Month Proclamation \\
Housing \& Homelessness & Fair Housing Month Proclamation; Housing Authority Consent Calendar; Homeless Solutions Update; Mobile Home Rent Stabilization \\
Public Safety & Illegal Fireworks Enforcement; Lenco Bearcat Armored Tactical Vehicle Purchase; Citywide Street Lighting and Safety; Police Misconduct Complaint \\
Economic Development & Small Business Saturday Proclamation; Economic Development Update; Sidewalk Vending Ordinance; Supervisor Hilda Solis Job Fair \\
Environmental Sustainability & Arbor Day Proclamation; Gas-Powered Leaf Blower Ban; Climate Emergency Declaration; Community Choice Energy (CCE) \\
Social Equity & Black History Month Proclamation; Women's History Month Proclamation; Veteran of the Year Recognitions; Support for Mayor and Council Against Harassment \\
Public Health & COVID-19 Pandemic Update; Breast Cancer Awareness Month Proclamation; Mental Health Awareness Month Proclamation; Mobile Vaccination for Homebound Residents \\
\bottomrule
\end{tabular}
\begin{minipage}{.85\textwidth}
\vspace{4pt}\footnotesize
\textit{Notes:} Example topics were generally selected from the most frequently occurring issues within each category. Additional examples were drawn to supplement top issues when the top issues were highly similar in content.
\end{minipage}
\end{table}

\begin{table}[H]
\centering
\caption{Public remote access duration}
\label{tab:acs_duration_ols}
\begin{tabular}{lc}
\toprule
 & Months of public remote access\\
\midrule
Log population & -0.776\\
 & (2.897)\\
Log median income & -10.607\\
 & (13.944)\\
Gini coefficient & -30.981\\
 & (78.819)\\
Renter share & 7.038\\
 & (24.568)\\
Hispanic share & -8.272\\
 & (18.028)\\
Black share & -21.402\\
 & (27.586)\\
Asian share & 4.758\\
 & (15.105)\\
BA+ share & 10.933\\
 & (30.589)\\
Racial diversity & -4.029\\
 & (22.491)\\
\midrule
Observations & 61\\
$R^2$ & 0.041\\
Mean dep. var. & 30.877\\
SD dep. var. & 12.202\\
\bottomrule
\end{tabular}
\begin{minipage}{0.85\textwidth}
\vspace{4pt}\footnotesize 
\textit{Notes:} This table reports an OLS regression of months of public remote access on 2019 ACS covariates for municipalities that ended live remote public comment and have matched ACS covariates. Standard errors in parentheses. $^{*} p<0.10$, $^{**} p<0.05$, $^{***} p<0.01$.
\end{minipage}
\begin{minipage}{.85\textwidth}
\vspace{4pt}\footnotesize
\textit{Notes:} Callaway-Sant\textquotesingle Anna (2021) estimator with not-yet-treated controls. Standard errors clustered at town level in parentheses. Treated towns split at the median of each dimension; all never-treated towns included as controls in both subsamples. Outcomes residualized on town$\times$month agenda issue shares within each subsample. $^{*}p<0.10$, $^{**}p<0.05$, $^{***}p<0.01$.
\end{minipage}
\end{table}

\begin{table}[!htbp]
\centering
\caption{Public remote access shutoff hazard}
\label{tab:hazard_zoom_shutoff}
\begin{tabular}{lc}
\toprule
 & Pr(public remote access shutoff in month $t$)\\
\midrule
Lagged county COVID cases & -0.000\\
 & (0.000)\\
Lagged vaccination rate & -0.020\\
 & (0.020)\\
Log population & -0.123\\
 & (0.279)\\
Log median income & -0.382\\
 & (0.532)\\
Gini coefficient & 5.724\\
 & (6.465)\\
Renter share & -3.622\\
 & (2.359)\\
Hispanic share & -1.974\\
 & (1.643)\\
Black share & 5.659**\\
 & (2.447)\\
Asian share & 1.511\\
 & (1.475)\\
BA+ share & -3.327\\
 & (2.172)\\
Racial diversity & -3.694*\\
 & (2.129)\\
\midrule
Observations & 3,022\\
Pseudo $R^2$ & 0.116\\
Mean dep. var. & 0.013\\
Time spline baseline & Yes\\
\bottomrule
\end{tabular}
\begin{minipage}{0.85\textwidth}
\vspace{4pt}\footnotesize \textit{Notes:} This table reports a logit discrete-time hazard model for the month-to-month probability of ending live public remote access. The specification includes one-month lags of county COVID cases and county vaccination rates, 2019 ACS covariates, and a spline in time since remote public participation adoption. Standard errors in parentheses. $^{*} p<0.10$, $^{**} p<0.05$, $^{***} p<0.01$.
\end{minipage}
\begin{minipage}{.85\textwidth}
\vspace{4pt}\footnotesize
\textit{Notes:} Callaway-Sant\textquotesingle Anna (2021) estimator with not-yet-treated controls. Standard errors clustered at town level in parentheses. Treated towns split at the median of each dimension; all never-treated towns included as controls in both subsamples. Outcomes residualized on town$\times$month agenda issue shares within each subsample. $^{*}p<0.10$, $^{**}p<0.05$, $^{***}p<0.01$.
\end{minipage}
\end{table}

\begin{table}[H]
\centering
\caption{Heterogeneity in Zoom shutoff effects, pre-treatment participation}
\label{tab:het_zoom_off_part}
\begin{tabular}{lcc}
\toprule
 & Above Median & Below Median \\
\midrule
NG Speakers & -4.943 & -0.457 \\
 & (3.809) & (1.372) \\
Share Female & -0.026 & 0.049 \\
 & (0.055) & (0.064) \\
Share Democrat & -0.026 & 0.007 \\
 & (0.054) & (0.064) \\
Mean Age & -0.613 & 2.488 \\
 & (1.549) & (3.252) \\
Share Below 40 & 0.027 & -0.100 \\
 & (0.032) & (0.084) \\
Share Above 65 & -0.012 & -0.010 \\
 & (0.048) & (0.066) \\
Age IQR & 0.105 & 1.112 \\
 & (1.872) & (3.638) \\
\bottomrule
\end{tabular}
\begin{minipage}{.85\textwidth}
\vspace{4pt}\footnotesize
\textit{Notes:} Callaway-Sant\textquotesingle Anna (2021) estimator with not-yet-treated controls. Standard errors clustered at town level in parentheses. Treated towns split at the median of each dimension; all never-treated towns included as controls in both subsamples. Outcomes residualized on town$\times$month agenda issue shares within each subsample. $^{*}p<0.10$, $^{**}p<0.05$, $^{***}p<0.01$.
\end{minipage}
\end{table}

\begin{table}[H]
\centering
\caption{Heterogeneity in remote access shutoff effects, pre-treatment speaker age}
\label{tab:het_zoom_off_age}
\begin{tabular}{lcc}
\toprule
 & Above Median & Below Median \\
\midrule
NG Speakers & -0.860 & -5.625 \\
 & (1.522) & (3.894) \\
Share Female & -0.043 & 0.063 \\
 & (0.054) & (0.059) \\
Share Democrat & -0.026 & -0.007 \\
 & (0.050) & (0.063) \\
Mean Age & -3.589*** & 5.945* \\
 & (1.006) & (3.127) \\
Share Below 40 & 0.080** & -0.162** \\
 & (0.031) & (0.080) \\
Share Above 65 & -0.102** & 0.089 \\
 & (0.045) & (0.064) \\
Age IQR & 1.199 & 0.127 \\
 & (1.902) & (3.730) \\
\bottomrule
\end{tabular}
\begin{minipage}{.85\textwidth}
\vspace{4pt}\footnotesize
\textit{Notes:} Callaway-Sant\textquotesingle Anna (2021) estimator with not-yet-treated controls. Standard errors clustered at town level in parentheses. Treated towns split at the median of each dimension; all never-treated towns included as controls in both subsamples. Outcomes residualized on town$\times$month agenda issue shares within each subsample. $^{*}p<0.10$, $^{**}p<0.05$, $^{***}p<0.01$.
\end{minipage}
\end{table}
 
\begin{table}[H]
\centering
\caption{Heterogeneity in remote access shutoff effects, log population}
\label{tab:het_zoom_off_pop}
\begin{tabular}{lcc}
\toprule
& Above Median & Below Median \\
\midrule
NG Speakers & -4.126 & -0.891 \\
 & (3.232) & (1.706) \\
Share Female & 0.007 & 0.008 \\
 & (0.061) & (0.055) \\
Share Democrat & 0.013 & -0.036 \\
 & (0.060) & (0.059) \\
Mean Age & 2.392 & -1.151 \\
 & (2.283) & (3.573) \\
Share Below 40 & -0.108 & 0.055 \\
 & (0.069) & (0.077) \\
Share Above 65 & 0.000 & -0.027 \\
 & (0.058) & (0.057) \\
Age IQR & -1.912 & 2.823 \\
 & (3.248) & (2.432) \\
\bottomrule
\end{tabular}
\begin{minipage}{.85\textwidth}
\vspace{4pt}\footnotesize
\textit{Notes:} Callaway-Sant\textquotesingle Anna (2021) estimator with not-yet-treated controls. Standard errors clustered at town level in parentheses. Treated towns split at the median of each dimension; all never-treated towns included as controls in both subsamples. Outcomes residualized on town$\times$month agenda issue shares within each subsample. $^{*}p<0.10$, $^{**}p<0.05$, $^{***}p<0.01$.
\end{minipage}
\end{table}

\begin{table}[H]
\centering
\caption{Heterogeneity in remote access shutoff effects, median income}
\label{tab:het_zoom_off_income}
\begin{tabular}{lcc}
\toprule
& Above Median & Below Median \\
\midrule
NG Speakers & -1.837 & -2.904 \\
 & (2.907) & (2.118) \\
Share Female & -0.041 & 0.052 \\
 & (0.058) & (0.054) \\
Share Democrat & -0.030 & 0.010 \\
 & (0.049) & (0.063) \\
Mean Age & -2.221* & 2.768 \\
 & (1.211) & (3.173) \\
Share Below 40 & 0.040 & -0.086 \\
 & (0.032) & (0.084) \\
Share Above 65 & -0.067 & 0.027 \\
 & (0.049) & (0.057) \\
Age IQR & 0.758 & -0.229 \\
 & (2.034) & (3.580) \\
\bottomrule
\end{tabular}
\begin{minipage}{.85\textwidth}
\vspace{4pt}\footnotesize
\textit{Notes:} Callaway-Sant\textquotesingle Anna (2021) estimator with not-yet-treated controls. Standard errors clustered at town level in parentheses. Treated towns split at the median of each dimension; all never-treated towns included as controls in both subsamples. Outcomes residualized on town$\times$month agenda issue shares within each subsample. $^{*}p<0.10$, $^{**}p<0.05$, $^{***}p<0.01$.
\end{minipage}
\end{table}

\begin{table}[H]
\centering
\caption{Heterogeneity in remote access shutoff effects, share BA+}
\label{tab:het_zoom_off_grad}
\begin{tabular}{lcc}
\toprule
& Above Median & Below Median \\
\midrule
NG Speakers & -4.144 & -0.605 \\
 & (2.632) & (2.383) \\
Share Female & -0.028 & 0.048 \\
 & (0.058) & (0.058) \\
Share Democrat & -0.075 & 0.067 \\
 & (0.049) & (0.063) \\
Mean Age & -2.423** & 3.722 \\
 & (1.232) & (3.205) \\
Share Below 40 & 0.065* & -0.132* \\
 & (0.038) & (0.079) \\
Share Above 65 & -0.053 & 0.020 \\
 & (0.050) & (0.063) \\
Age IQR & 1.161 & -0.196 \\
 & (2.118) & (3.686) \\
\bottomrule
\end{tabular}
\begin{minipage}{.85\textwidth}
\vspace{4pt}\footnotesize
\textit{Notes:} Callaway-Sant\textquotesingle Anna (2021) estimator with not-yet-treated controls. Standard errors clustered at town level in parentheses. Treated towns split at the median of each dimension; all never-treated towns included as controls in both subsamples. Outcomes residualized on town$\times$month agenda issue shares within each subsample. $^{*}p<0.10$, $^{**}p<0.05$, $^{***}p<0.01$.
\end{minipage}
\end{table}

\begin{table}[H]
\centering
\caption{Heterogeneity in remote access shutoff effects, share renter}
\label{tab:het_zoom_off_rent}
\begin{tabular}{lcc}
\toprule
& Above Median & Below Median \\
\midrule
NG Speakers & -3.728 & -1.649 \\
 & (3.477) & (2.583) \\
Share Female & 0.031 & -0.015 \\
 & (0.061) & (0.053) \\
Share Democrat & 0.046 & -0.059 \\
 & (0.068) & (0.042) \\
Mean Age & 4.972 & -4.182*** \\
 & (3.027) & (0.982) \\
Share Below 40 & -0.129 & 0.082** \\
 & (0.080) & (0.033) \\
Share Above 65 & 0.084 & -0.127*** \\
 & (0.054) & (0.044) \\
Age IQR & 1.278 & -0.347 \\
 & (3.716) & (1.819) \\
\bottomrule
\end{tabular}
\begin{minipage}{.85\textwidth}
\vspace{4pt}\footnotesize
\textit{Notes:} Callaway-Sant\textquotesingle Anna (2021) estimator with not-yet-treated controls. Standard errors clustered at town level in parentheses. Treated towns split at the median of each dimension; all never-treated towns included as controls in both subsamples. Outcomes residualized on town$\times$month agenda issue shares within each subsample. $^{*}p<0.10$, $^{**}p<0.05$, $^{***}p<0.01$.
\end{minipage}
\end{table}

\begin{table}[H]
\centering
\caption{Heterogeneity in remote access shutoff effects, share White non-Hispanic}
\label{tab:het_zoom_off_white}
\begin{tabular}{lcc}
\toprule
& Above Median & Below Median \\
\midrule
NG Speakers & -5.677* & -0.261 \\
 & (3.321) & (2.391) \\
Share Female & -0.031 & 0.066 \\
 & (0.052) & (0.064) \\
Share Democrat & -0.007 & -0.019 \\
 & (0.047) & (0.067) \\
Mean Age & -2.389 & 4.736 \\
 & (1.464) & (3.158) \\
Share Below 40 & 0.048 & -0.117 \\
 & (0.039) & (0.079) \\
Share Above 65 & -0.040 & 0.031 \\
 & (0.053) & (0.065) \\
Age IQR & -0.269 & 1.283 \\
 & (1.951) & (3.557) \\
\bottomrule
\end{tabular}
\begin{minipage}{.85\textwidth}
\vspace{4pt}\footnotesize
\textit{Notes:} Callaway-Sant\textquotesingle Anna (2021) estimator with not-yet-treated controls. Standard errors clustered at town level in parentheses. Treated towns split at the median of each dimension; all never-treated towns included as controls in both subsamples. Outcomes residualized on town$\times$month agenda issue shares within each subsample. $^{*}p<0.10$, $^{**}p<0.05$, $^{***}p<0.01$.
\end{minipage}
\end{table}

\begin{table}[H]
\centering
\caption{Heterogeneity in remote access shutoff effects, share Black}
\label{tab:het_zoom_off_black}
\begin{tabular}{lcc}
\toprule
& Above Median & Below Median \\
\midrule
NG Speakers & -0.550 & -5.047 \\
 & (2.468) & (3.299) \\
Share Female & -0.016 & 0.063 \\
 & (0.064) & (0.050) \\
Share Democrat & 0.003 & -0.027 \\
 & (0.047) & (0.072) \\
Mean Age & 0.103 & 2.577 \\
 & (3.192) & (2.097) \\
Share Below 40 & -0.003 & -0.072 \\
 & (0.072) & (0.065) \\
Share Above 65 & -0.026 & 0.021 \\
 & (0.063) & (0.060) \\
Age IQR & 4.494* & -3.343 \\
 & (2.710) & (3.177) \\
\bottomrule
\end{tabular}
\begin{minipage}{.85\textwidth}
\vspace{4pt}\footnotesize
\textit{Notes:} Callaway-Sant\textquotesingle Anna (2021) estimator with not-yet-treated controls. Standard errors clustered at town level in parentheses. Treated towns split at the median of each dimension; all never-treated towns included as controls in both subsamples. Outcomes residualized on town$\times$month agenda issue shares within each subsample. $^{*}p<0.10$, $^{**}p<0.05$, $^{***}p<0.01$.
\end{minipage}
\end{table}

\end{document}

%% file: _header.tex
\usepackage[utf8]{inputenc}
\usepackage[english]{babel}
\usepackage{amsfonts}
\usepackage{siunitx}
\usepackage{amssymb}
\usepackage{enumitem}
\usepackage{mathtools}
\usepackage{comment}
\usepackage{xcolor}
\usepackage{caption}
\usepackage{subcaption}
\usepackage{booktabs}	
\usepackage{setspace}
\usepackage{longtable}
\usepackage{graphicx}
\usepackage{fullpage}
\usepackage{soul}
\usepackage{placeins}
\usepackage{tcolorbox}
\usepackage[backref=page,breaklinks=true,hidelinks]{hyperref}
\hypersetup{
    colorlinks,
    linkcolor={red!50!black},
    citecolor={blue!50!black},
    urlcolor={blue!80!black}
}
\renewcommand*{\backref}[1]{}
\renewcommand*{\backrefalt}[4]{%
    \ifcase #1 {\footnotesize(Not cited.)}%
    \or        {\footnotesize(Cited on page~#2.)}%
    \else      {\footnotesize(Cited on pages~#2.)}%
    \fi}
\usepackage{amsthm}
\usepackage{thmtools}
\usepackage{dsfont}
\usepackage{natbib}
\usepackage{mleftright}
\usepackage{xfrac}
\usepackage{rotating}
\usepackage{mdframed}

\presetkeys%
    {todonotes}%
    {inline,backgroundcolor=yellow}{}

\usepackage{fvextra}
\DefineVerbatimEnvironment{Verbatim}{Verbatim}{breaklines=true}

\newcommand\nnfootnote[1]{%
  \begin{NoHyper}
  \renewcommand\thefootnote{}\footnote{#1}%
  \addtocounter{footnote}{-1}%
  \end{NoHyper}
}